\documentclass[twocolumn,preprintnumbers,amsmath,prb,amssymb]{revtex4-1}
\usepackage {amssymb,amsmath}
\usepackage {graphicx}
\usepackage{subfigure}
\usepackage[usenames]{color}
\def \be {\begin{equation}}
\def \ee {\end{equation}}
\def \ben {\begin{eqnarray}}
\def \een {\end{eqnarray}}
\def\re#1{{ \color{black} {#1}}}
\usepackage{multirow}
\usepackage{ctable}

\begin {document}       
\bibliographystyle{prsty}           
  \title{Theoretical investigation of non-F\"{o}rster exciton transfer mechanisms in perylene diimide donor, phenylene bridge, and terrylene diimide acceptor systems}

  \author{Lei Yang}
  \affiliation{Center for Molecular Systems and Organic Devices, Key Laboratory for Organic Electronics and Information Displays and Jiangsu Key Laboratory for Biosensors, Institute of Advanced Materials, Nanjing University of Posts and Telecommunications, 9 Wenyuan Road, Nanjing 210023, China}
  \author{Seogjoo J. Jang\footnote{Corresponding Author, Email:SJANG@qc.cuny.edu}}
 \affiliation{Department of Chemistry and Biochemistry, Queens College, City University of New York, 65-30 Kissena Boulevard, Queens, New York 11367\footnote{Primary Affiliation and Mailing Address} \& PhD programs in Chemistry and Physics, and Initiative for the Theoretical Sciences, Graduate Center, City University of New York, 365 Fifth Avenue, New York, NY 10016}

  \date{Published in {\it the Journal of Chemical Physics}\footnote{Invited contribution to the Special Topic on Excitons: Energetics and Spatiotemporal Dynamics} {\bf 153}, 144305 (2020)}

 \begin{abstract}
\re{The rates of exciton transfer within dyads of perylene diimide  and terrylene diimide connected by oligophenylene bridge units have been shown to deviate significantly from those of F\"{o}rster's resonance energy transfer theory, according to single molecule spectroscopy experiments.  The present work provides}  a detailed computational and theoretical study investigating the source of such discrepancy.  Electronic spectroscopy data are calculated by time-dependent density function theory  and then compared with experimental results. Electronic couplings \re{between exciton donor and acceptor} are estimated based on both transition density cube method and transition dipole approximation.
These results confirm that the delocalization of exciton to the bridge parts contribute to \re{significant} enhancement of donor-acceptor electronic coupling.  Mechanistic details of \re{exciton transfer}  are examined by estimating the contributions of the bridge electronic states, vibrational modes of the dyads commonly coupled to both donor and acceptor, inelastic resonance energy transfer mechanism, and dark exciton states. These analyses suggest that the contribution of common vibrational modes serves as the main source of deviation from F\"{o}rster's spectral overlap expression.
 \end{abstract}

\maketitle

\section{Introduction}
F\"{o}rster's resonance energy transfer (FRET) theory\cite{forster-ap,forster-dfs}  is a particular form of Fermi's golden rule applied to the transfer of localized excitons through transition dipole interactions.  As the first \re{quantum mechanical} theory prescribing calculation of \re{exciton transfer rate} from experimentally measurable quantities, the FRET theory \re{for decades} has played a central role in elucidating numerous luminescence processes\cite{silbey-arpc27,ret,nitzan,scholes-arpc54,olaya-castro-irpc30,jang-wires3,bardeen-arpc65,haacke-burghardt,jang-rmp90} and for the determination of nanometer scale distances.\cite{stryer-pnas58,roy-nm5,selvin-nsb7,sahoo-jppc12,heyduk-cob13,schuler-cosb18,ha-pnas93,roy-nm5,weiss-science283,guo-pccp16,basak-pccp16,chung-pccp16,stockmar-jpcb120}  Nonetheless, it is not unreasonable to expect many exciton transfer processes going beyond the FRET mechanism, \re{considering simple assumptions behind the theory}.   Indeed, there have been various experimental results\cite{beljonne-jpcb113,schuler-pnas102,langhals-jacs132,dayal-jacs128,metivier-prl98,hinze-jcp128,athanasopoulos-jpcl8} indicating non-F\"{o}rster mechanisms, which are too many to cite here all.  There have also been theories of exciton transfer/dynamics\cite{sumi-prl50,sumi-jpcb103,jang-prl92,jang-cp275,jang-jcp127,jang-jcp129,jang-jcp131,jang-jcp135,yang-jcp137,jang-wires,jang-prl113,jang-jpcc123,hennebicq-jcp130,du-cs9,jang-exciton} for more general cases than the FRET theory.  However, actual examples of direct and clear verification of non-FRET processes have been surprisingly few.  One exceptional example in this regard is the non-FRET process observed for the transfer of excitons within dyads consisting of perylene diimide (PDI) and terrylene diimide (TDI) molecules with end-standing phenyl rings, {\it i.e.}, Ph-PDI and TDI-Ph.\cite{metivier-prl98,fuckel-jcp128,hinze-jcp128,curutchet-jpcb112}  

\begin{figure}
\vspace{.2in}
\begin{tabular}{c}
\includegraphics[scale=0.4]{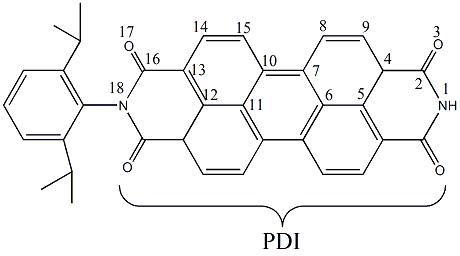} \\
Ph-PDI    \\
\includegraphics[scale=0.4]{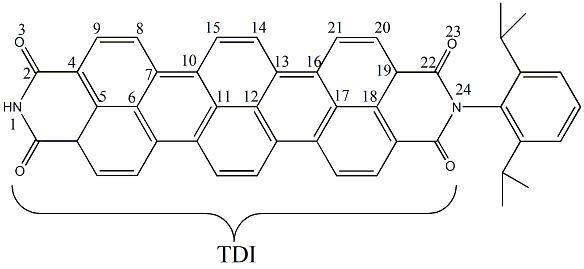} \\
TDI-Ph    \\
\end{tabular}
\caption{Sketch maps of Ph-PDI and TDI-Ph.}
\label{Sketch-map}
\end{figure}

 A series of experimental and computational studies have been made for two types of dyads, where Ph-PDI and TDI-Ph (see Fig.~\ref{Sketch-map} for sketch structures of Ph-PDI and TDI-Ph) are linked by two types of oligophenylene spacers, one linear p-terphenyl bridge and the other hepta(phenylene) bridge containing a kink. 
We denote them collectively as Ph-PDI-nPh-TDI-Ph, and call the one with linear p-terphenyl bridge ($n=3$) as Dyad 1 ({\bf D1}) and the one with  hepta(phenylene) bridge ($n=7$) as Dyad 2 ({\bf D2}).  A number of experimental measurements and theoretical studies\cite{hubner-jcp120,hinze-jpca109,fuckel-jcp125,metivier-prl98,hinze-jcp128,curutchet-jpcb112,fuckel-jcp128} on these systems have revealed evidence for non-FRET behavior.  Namely, the experimentally determined RET rates for these systems were found to be faster than the estimates of FRET rates, and the experimental data for {\bf D1} \re{in particular} could not be fully accounted for by F\"{o}rster's spectral overlap expression.

The extents of non-FRET mechanisms for {\bf D1} and {\bf D2} are somewhat different.  Basch\'{e} and coworkers\cite{metivier-prl98,hinze-jcp128} determined RET rates for {\bf D1} at $1.4\ {\rm K}$ by measuring line widths of single molecule fluorescence excitation spectra, and reported \re{the average experimental rate to be} $3.2\times 10^{11} {\rm s^{-1}}$.  For {\bf D2} at room temperature, they employed single photon counting techniques,\cite{hinze-jcp128} and determined the average experimental rate to be $1.2\times 10^{10}\ {\rm\ s^{-1}}$.  On the other hand, the average value of theoretical FRET rates for {\bf D1} at the low temperature limit was estimated to be $3.9-4.0\times 10^{10}\ {\rm\ s^{-1}}$, and that for {\bf D2} at room temperature was estimated to be $4.9-7.3 \times 10^{9}\ {\rm s^{-1}}$.  Thus, the discrepancy between experimental results and FRET estimates are about a factor of 10 for {\bf D1} and about a factor of 2 for {\bf D2}. In addition, low temperature single molecule experiments for {\bf D1} revealed that the measured RET rates are not proportional to the spectral overlap between the donor and the acceptor,\cite{hinze-jcp128,metivier-prl98} strongly indicating the presence of additional mechanisms that are not accounted for by the FRET theory.

The reported deviations from the FRET theory may be explained by a number of  structural or environmental features of the molecules.
First,  both bridge linkers in the dyads are not entirely rigid and can alter the distance between the chromophores as well as their relative orientations.  Indeed, Hubner {\it et al.} observed\cite{hubner-jcp120} significant deviation from collinearity for {\bf D1}, with the distribution of misalignment angles peaking around 22$^{\circ}$. Subsequently, Fuckel $et~al.$ used a Monte Carlo simulation to show that the p-phenylene oligomer has a rather high flexibility.\cite{fuckel-jcp125}
\re{However,} these results suggest much more flexibility in the longer hepta(phenylene) bridge \re{than the shorter one and cannot explain why non-F\"{o}rster mechanism is more significant in {\bf D1} than in {\bf D2}}.
Second, the polarizability of the bridge unit induces a slight delocalization of PDI and TDI transition densities.  This can enhance the effective donor-acceptor electronic coupling, and thus the rate of RET.\cite{fuckel-jcp128,curutchet-jpcb112}
On the other hand, Hinze $et~al.$\cite{hinze-jcp128} and M$\acute{e}$tivier $et~al.$\cite{metivier-prl98} have also shown that the static disorder and the flexibility of the dyads can make substantial contributions at both room and low temperature. However, possible effects of common vibrational modes, inelastic processes, or multichromophoric effects have not yet been examined for these systems.

For {\bf D2}, quantitative account of experimental RET rates seem possible \re{considering both} the delocalization of excitons to the bridge unit and \re{conformational fluctuations of molecules}.\cite{fuckel-jcp128}   However, for {\bf D1}, \re{even after considering such effects,} there was discrepancy of about a factor of 2 between experimental rates and the best theoretical estimates based on the FRET theory.\cite{fuckel-jcp128,curutchet-jpcb112} Understanding the major source of this discrepancy still remains an important theoretical issue. 

In the present paper, we make careful theoretical investigation of potential factors contributing to the non-FRET mechanisms in the two dyads, {\bf D1} and {\bf D2}. 
In Sec. II, the extent of the delocalization of transition densities are reexamined in order to obtain more definite information on the contributions of the through-bond interaction to the non-FRET behavior.  Section III presents three plausible non-FRET mechanisms, and provides quantitative explanation of the fast RET rate for {\bf D1}.   Section IV offers the conclusion of this work.

\begin{figure}
\begin{tabular}{cc}
(a)& \\
&\includegraphics[width=3.2 in]{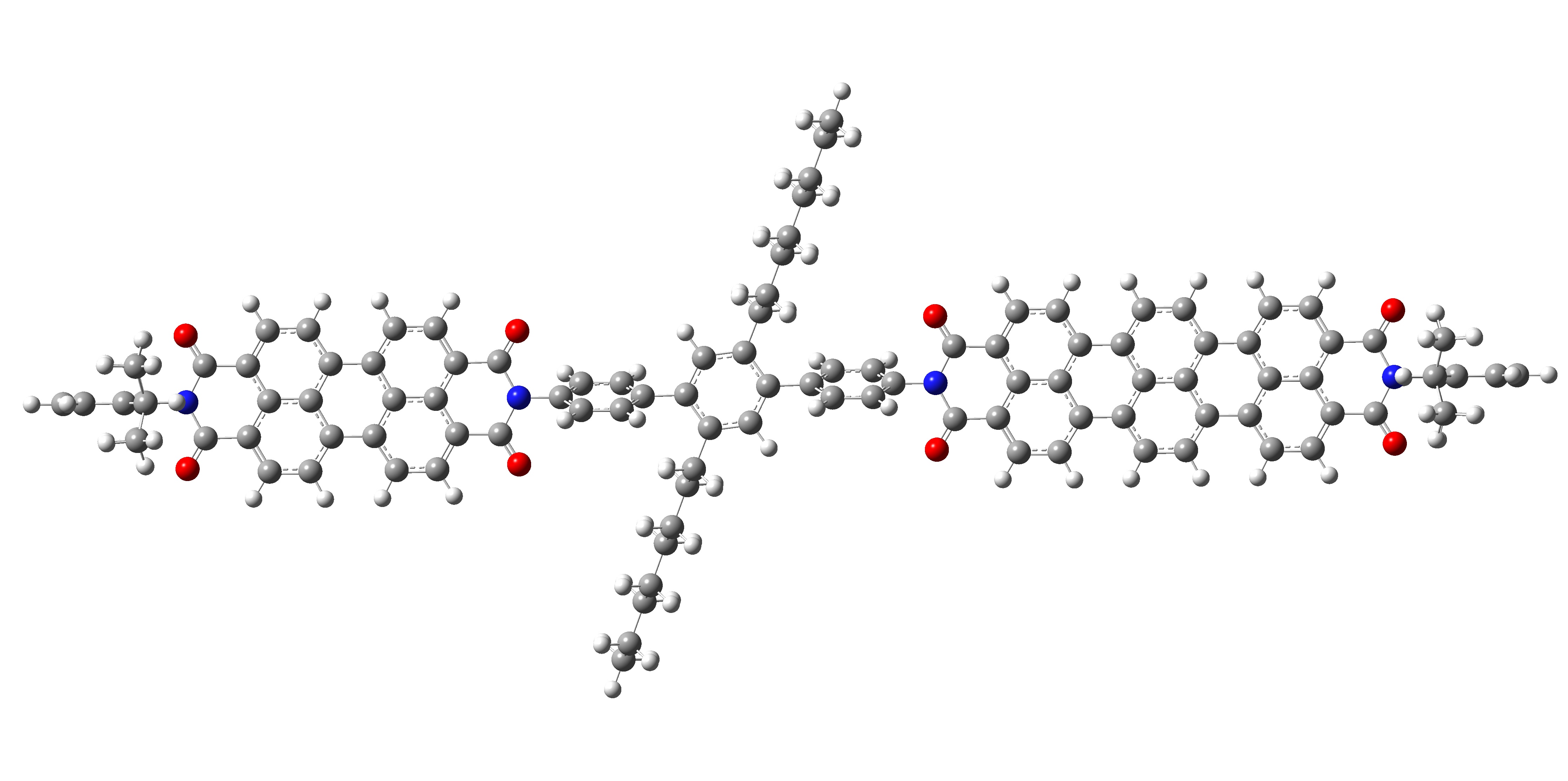} \\
(b)&\\
& \includegraphics[width=3.2 in]{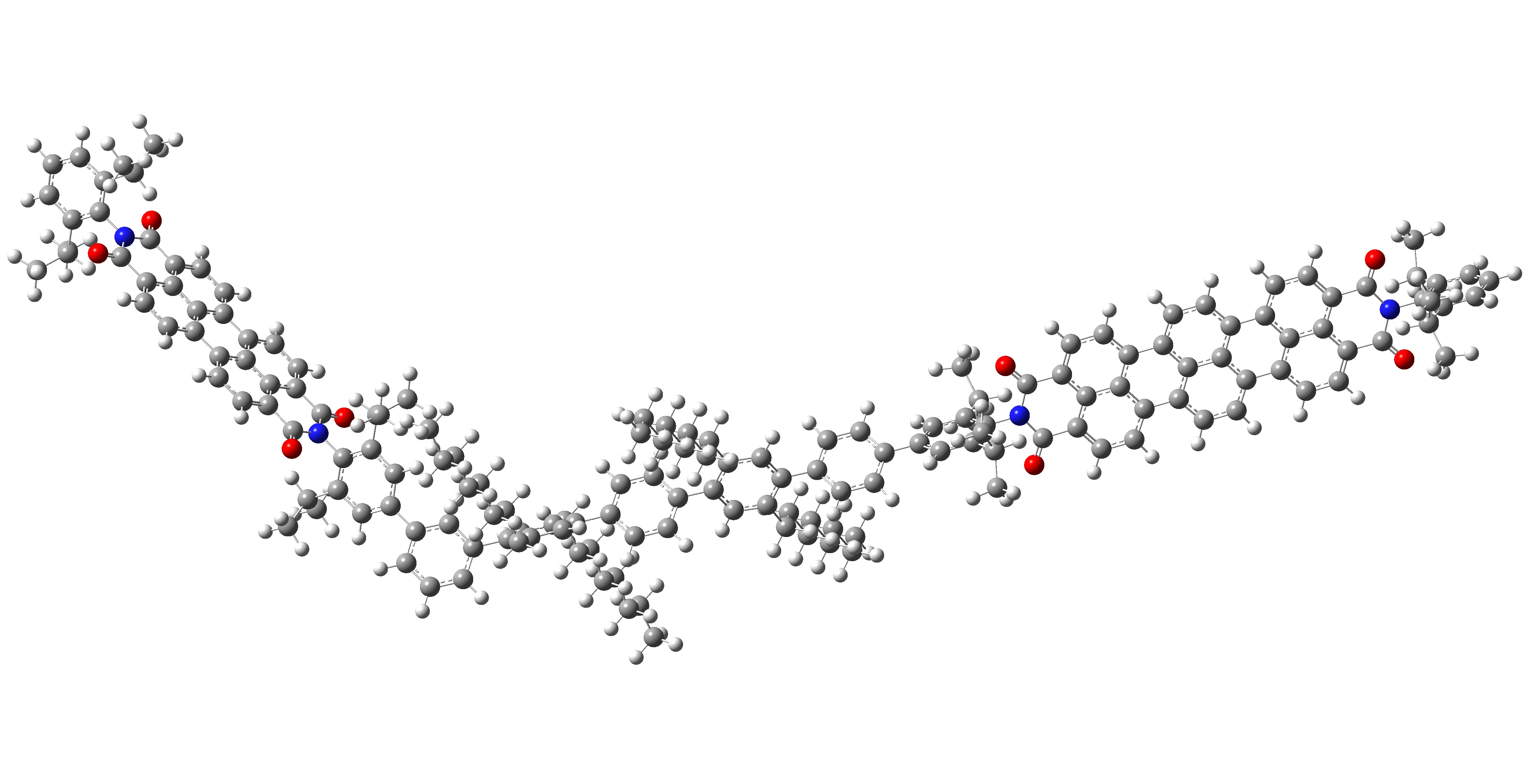} 
\end{tabular}
\caption{Optimized molecular structures of (a) Ph-PDI-3Ph-TDI-Ph and  (b) Ph-PDI-7Ph-TDI-Ph at the B3LYP/6-31G(d) level.}
\label{Actual-structure}
\end{figure}

\section{Computational study}

\subsection{Electronic Structure Calculation and Optimization}
   This section provides results of electronic structure calculation for the dyads and their units. For the ground states of the molecules, structures were optimized by the density functional theory (DFT) method.\cite{kohn-pr140} While we used Becke's three-parameter hybrid exchange functional and Lee-Yang-Parr gradient-corrected correlation function (B3LYP functional)\cite{lee-prb37} as the main functional, CAM-B3LYP functional\cite{yanai-cpl393} and M06-2X functional\cite{zhao-tca120} were also used for select cases for more reliable computational assessment.  The time-dependent density functional theory (TD-DFT) method\cite{runge-prl52} was employed to calculate excitation energies and for the optimization of excited states, employing B3LYP functional for all cases and CAM-B3LYP and M06-2X functionals for monomers.  In addition,  excited state structures optimized by configuration interaction singles (CIS)\cite{foresman-jpc96} were also used for the case where the B3LYP functional is used for the TD-DFT calculation of emission transition energies.   All the calculations were made employing the 6-31G(d) basis set except for the structural optimization of {\bf D2} by CIS, for which 3-21G basis set was used.
All of the above calculations were carried out with the Gaussian 09 package.\cite{g09}
\re{The ground state structures of the dyads optimized by DFT with B3LYP functional are shown in Fig.~\ref{Actual-structure}. 
 Detailed information on structures of these and monomer units are provided in Appendix A.}  
  
\begin{table}

\caption{\label{pdi-ph-ex} Calculated excitation energies (in ${\rm eV}$) \re{and oscillator strength ($f$)} for Ph-PDI employing TD-DFT with three different functionals and 6-31G(d) basis.  Experimental data\cite{hinze-jcp128} are also listed for comparison. Computational data for ${\rm S_0\rightarrow S_1}$ were calculated for optimized structures for the ground electronic state (${\rm S_0}$), and those for the ${\rm S_1\rightarrow S_0}$ were calculated for optimized structures for the first excited singlet state (${\rm S_1}$).  Both vacuum (v.) and toluene solvent (s.) modeled at the level of PCM are provided.\vspace{.1in}}

\begin{tabular}{c|c|l|c|c|c}
\hline
\hline
Transitions & Method      &env.         &$f$ &${\rm E_{cal}  (eV)}$           &${\rm E_{exp} (eV)}$   \\
        \hline
\multirow{6}*{${\rm S_{0}\rightarrow S_{1}}$} &\multirow{2}*{B3LYP}    &v.   &   0.6196 & 2.4275                 &\multirow{6}*{2.3706}  \\

                                                               &                                      &s. &0.9196 &2.3018  &                     \\
\cline{2-5}
          &\multirow{2}*{CAM-B3LYP}      &v.  &0.8794 &2.8349&\\     
          
         &      &s. &  1.0751  & 2.6954                               \\ 
         
\cline{2-5}         
           &\multirow{2}*{M06-2X}          &v.   & 0.8646 &2.8494      \\

  & &s.& 1.0558 & 2.7137 &\\
       \hline      
 \multirow{7}*{${\rm S_{1}\rightarrow S_{0}}$}&\multirow{2}*{B3LYP}    &v.\footnote{Transition from $S_2$}   &0.7259&2.2338  & \multirow{7}*{2.3393} \\

  & &s. &0.9089 &2.0518     &                   \\
 \cline{2-5} 
 &B3LYP/CIS & v. & 0.7021 & 2.2638  &   \\
 \cline{2-5}
            &\multirow{2}*{CAM-B3LYP} &v.&0.8537&2.4441&                                  \\
         
          &&s. & 1.0527& 2.2632  &\\ 
\cline{2-5}
                    &\multirow{2}*{M06-2X}  &v.   &0.8362&2.4615   &                     \\
                   
                   &&s. &1.0335 &2.2843       \\
                   \hline
       
   & \multirow{2}*{B3LYP}  &       v. &    & 2.0722                &\multirow{6}*{2.3184}             \\
 
 &&s. &  &2.1634  &\\
 \cline{2-5}
$0-0$       &\multirow{2}*{CAM-B3LYP} &v.& &2.6295  & \\
  
 $({\rm S_0-S_1})$ &&s. & & 2.4653 & \\
  \cline{2-5}
   &\multirow{2}*{M06-2X} &v. & & 2.6531& \\
   
   &&s.& & 2.4818& \\
       \hline
       \hline
         \end{tabular}
\end{table}

\begin{table}
\caption{\label{tdi-ph-ex} Calculated excitation energies (in ${\rm eV}$) for TDI-Ph.  Detailed methods and conventions of notations are the same as those in Table  \ref{pdi-ph-ex}.  \vspace{.1in}}

 \begin{tabular}{c|c|l|c|c|c}

\hline
\hline
Transitions & Method      &env.         &$f$  &${\rm E_{cal}  (eV)}$          &${\rm E_{exp} (eV)}$   \\
        \hline
\multirow{6}*{${\rm S_{0}\rightarrow S_{1}}$} &\multirow{2}*{B3LYP}    &v. &  1.2072  & 1.9668                 &\multirow{6}*{1.9075}  \\

                                                               &                                      &s. & 1.4606 &1.8097    &                     \\
\cline{2-5}
          &\multirow{2}*{CAM-B3LYP}      &v.  &1.3963 &2.3540&\\     
          
                                                      &      &s. &1.6444&2.1951 & \\ 
         
\cline{2-5}         
           &\multirow{2}*{M06-2X}          &v.    & 1.3711 &2.3487     \\

                                                       & &s. &1.6159 &2.1932   &\\
       \hline      
 \multirow{7}*{${\rm S_{1}\rightarrow S_{0}}$}&\multirow{2}*{B3LYP}    &v.  &1.1788 & 1.7992 & \multirow{7}*{1.8786} \\
 
                                                                                                     & &s. &1.4446 &1.6146    &                   \\
 \cline{2-5} 
 &B3LYP/CIS & v. & 1.1677 & 1.8562 & \\
 \cline{2-5} 
            &\multirow{2}*{CAM-B3LYP} &v. &1.3626   &2.0079 &                                \\
           
                                                           &&s. &1.6180 &1.7966  &\\ 
\cline{2-5}
                    &\multirow{2}*{M06-2X}  &v.  &1.3409  &2.0144  &                     \\
                  
                                                          &&s. &1.5944 &1.8086       \\
                   \hline
       
   & \multirow{2}*{B3LYP}       &  v.    & & 1.8808              &\multirow{6}*{1.8846}             \\
 
 &&s. & &1.7037   &\\
 \cline{2-5}
$0-0$       &\multirow{2}*{CAM-B3LYP} &v.& &2.1680   & \\
  
 $({\rm S_0-S_1})$ &&s. &&1.9849 & \\
  \cline{2-5}
   &\multirow{2}*{M06-2X} &v. && 2.1757 & \\
   
   &&s.&&1.9928 & \\
       \hline
       \hline
        \end{tabular}
\end{table}

Tables~\ref{pdi-ph-ex} and \ref{tdi-ph-ex} provide electronic transition energies and oscillator strengths for Ph-PDI and TDI-Ph, respectively, which were calculated by the TD-DFT method with three different functionals.  Relevant experimental data\cite{hubner-jcp120,fuckel-jcp125,metivier-prl98,hinze-jcp128} are provided as well.  Transition energies including solvation effect calculated at the level of polarizable continuum model (PCM) for toluene, are provided as well.   

For the case of Ph-PDI shown in Table \ref{pdi-ph-ex}, the transition energies from the ground state to the first excited state calculated by B3LYP functional in vacuum are shown to be very close to the experimental value and are comparable to previous calculations based on higher level methods or larger basis set.\cite{fuckel-jcp128,curutchet-jpcb112} On the other hand, those based on CAM-B3LYP and M06-2X significantly overestimate these values.  However, excited state calculation employing the B3LYP functional is shown to produce unreliable results.  First of all, excited state optimization employing the B3LYP functional leads to a dark first excited state. This is in contrast to the results for PDI only (without end Ph unit) as reported in Ref. \onlinecite{fuckel-jcp128}.   In addition, the emission transition energy from the second excited state (for a structure optimized for ${\rm S_1}$)  significantly underestimates the experimental emission energy.   Inclusion of solvation effect further reduces the transition energy.  In contrast, both calculations employing CAM-B3LYP and M06-2X functionals result in bright first excited states and produce transition energies very close to the experimental value. When CIS is used for the optimization of the excited state structure, we find that the deficiency of B3LYP can be corrected to some extent.  Interestingly, the resulting transition energy ($2.2638\ {\rm eV}$) in vacuum is very close to those calculated by CAM-B3LYP and M06-2X functional including solvation effect.   While this appears to be coincidence due to cancellation of errors, it can be useful in practice for efficient estimation of the emission energy. 

As can be seen from Table \ref{tdi-ph-ex}, the trends of calculation results based on different functionals for TDI-Ph are similar to those of Ph-PDI.   For the absorption energy  from the ground to the first excited state, optimization and calculation of excitation energy using B3LYP functional in vacuum provides fairly accurate estimate.  For the emission energy from the first excited state, optimization by CIS and calculation of the transition energy employing B3LYP in vacuum again result in a value very close to the experimental one.  Thus, we employ these methods for the calculation of electronic transition energies of {\bf D1} and {\bf D2}.

Table~\ref{Excitation-energy} provides absorption energies for both dyads calculated by B3LYP with structures optimized by B3LYP, and emission energies calculated by B3LYP with structures optimized by CIS.   All the calculated results are very close to experimental ones.  This is consistent with the trends shown in Tables \ref{pdi-ph-ex} and \ref{tdi-ph-ex}.  These calculation results also help more detailed assignment of transitions corresponding to experimental peaks.   By comparing these with relative energies of frontier molecular orbitals, \re{as detailed in Appendix A}, we find that the excitations from the ground states of both dyads can be related to combinations of individual transitions for Ph-PDI and TDI-Ph. For example, ${\rm S_{0}\rightarrow S_{2}}$ and ${\rm S_{0}\rightarrow S_{6}}$ transitions of {\bf D1} are very close to ${\rm S_{0}\rightarrow S_{1}}$ transitions of TDI-Ph and Ph-PDI, respectively.

For emission spectra of both dyads, the trends are somewhat different. The emission transition energies for TDI parts are close to those of TDI-Ph. On the other hand, those of transitions localized at the PDI parts are quite different from those of the emission spectra of Ph-PDI. This trend is consistent with the observation that the bond lengths in PDI parts change significantly after the excitation in both dyads.  The transition energies of both dyads are red-shifted when compared with the individual parts. This indicates that the mediating phenyl unit makes some contribution to the electronic excitation.  We will examine this issue in more detail below.

\begin{table}
\caption{\label{Excitation-energy} Excitation energies and oscillator strengths calculated for two dyads, and related experimental values.\cite{hinze-jcp128}  For transitions from ${\rm S_0}$, structures were optimized by DFT for the ground state and excitation energies were calculated by TD-DFT.  For the transitions from the excited states, structures were optimized for ${\rm S_1}$ by CIS and energies were calculated by TD-DFT.  6-31(d) basis was used for all the calculations except for for the structural optimization of {\bf D2}. \vspace{.1in}}

       \begin{tabular}{c|c|c|c|c|c}
\hline
\hline
Dyad  & Method               &transition         &$f$      &${\rm E_{cal}(eV)} $    &${\rm E_{exp} (eV)}$       \\
        \hline
\multirow{4}*{1}  &\multirow{2}*{B3LYP}&${\rm S_{0}\rightarrow S_{2}}$ &1.5413  &1.9506    &1.9075                      \\
        
                     &  &${\rm S_{0}\rightarrow S_{6}}$ &0.8771 &2.4150     &2.3706                     \\
        \cline{2-6}
                   &\multirow{2}*{B3LYP/CIS}       &${\rm S_{1}\rightarrow S_{0}}$  &1.4380  &1.8411     &1.8786                   \\
        
                    &       &${\rm S_{6}\rightarrow S_{0}}$ &0.8993 &2.6159     &2.3393                    \\
        \hline
\multirow{4}*{2}  &\multirow{2}*{B3LYP}&${\rm S_{0}\rightarrow S_{2}}$ &1.5310 &1.9455     &1.9075                     \\
        
                     &      &${\rm S_{0}\rightarrow S_{8}}$ &0.9155 &2.4056     &2.3706                    \\
        \cline{2-6}
                   &\multirow{2}*{B3LYP/CIS\footnote{3-21G basis for CIS}}       &${\rm S_{1}\rightarrow S_{0}}$ &1.4794 &1.8196     &1.8786                       \\
        
                    &          &${\rm S_{14}\rightarrow S_{0}}$ &0.9259 &2.5783    &2.3393                     \\
        \hline
        \hline
        \end{tabular}
\end{table}

\begin{figure}
\begin{tabular}{|c|c|}
\hline
\includegraphics[scale=0.042]{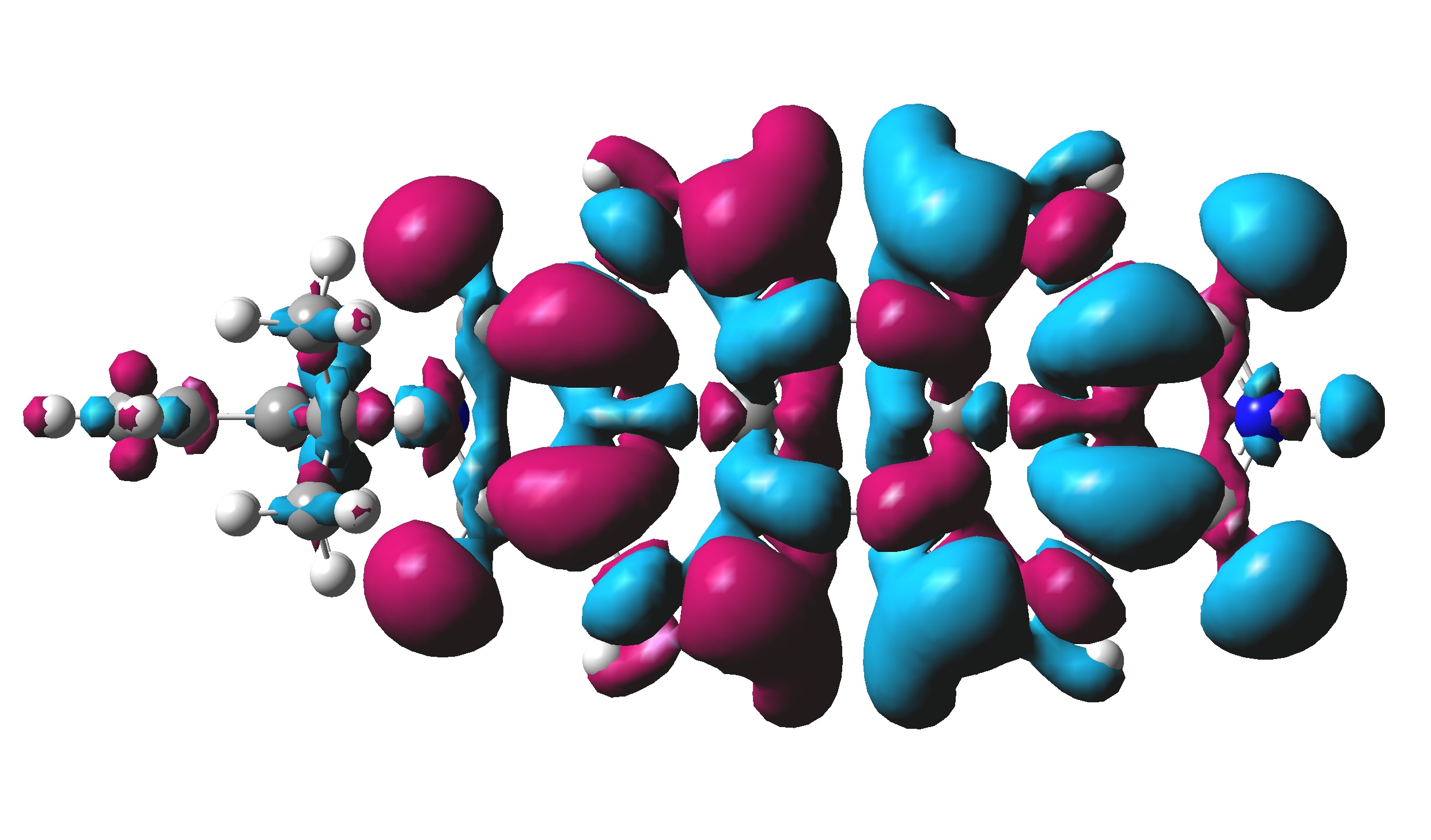} &
\includegraphics[scale=0.042]{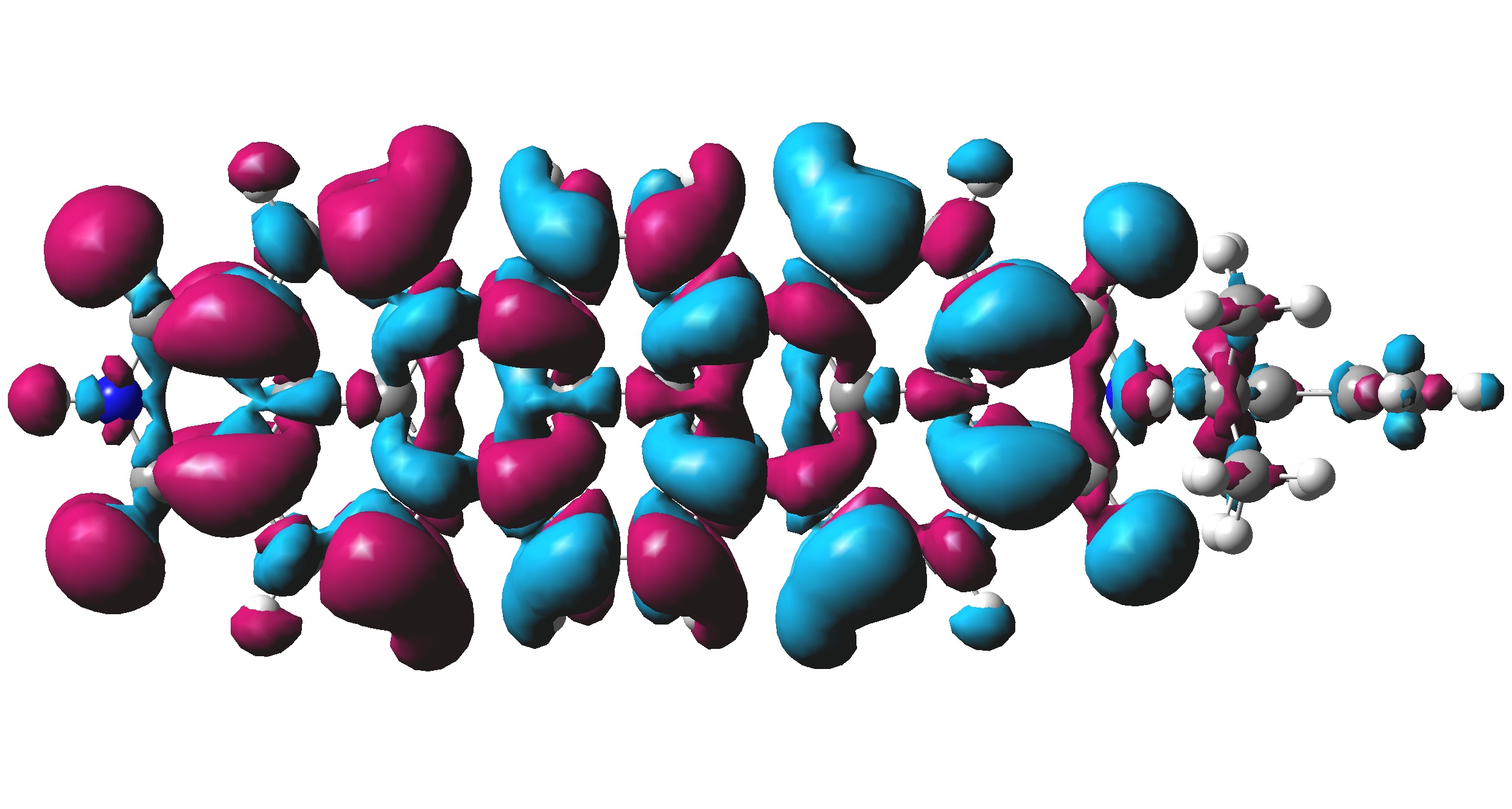} \\
(a) Ph-PDI (${\rm S_1\rightarrow S_0}$)   &  (b)TDI-Ph (${\rm S_0\rightarrow S_1}$)          \\
\hline
\includegraphics[scale=0.038]{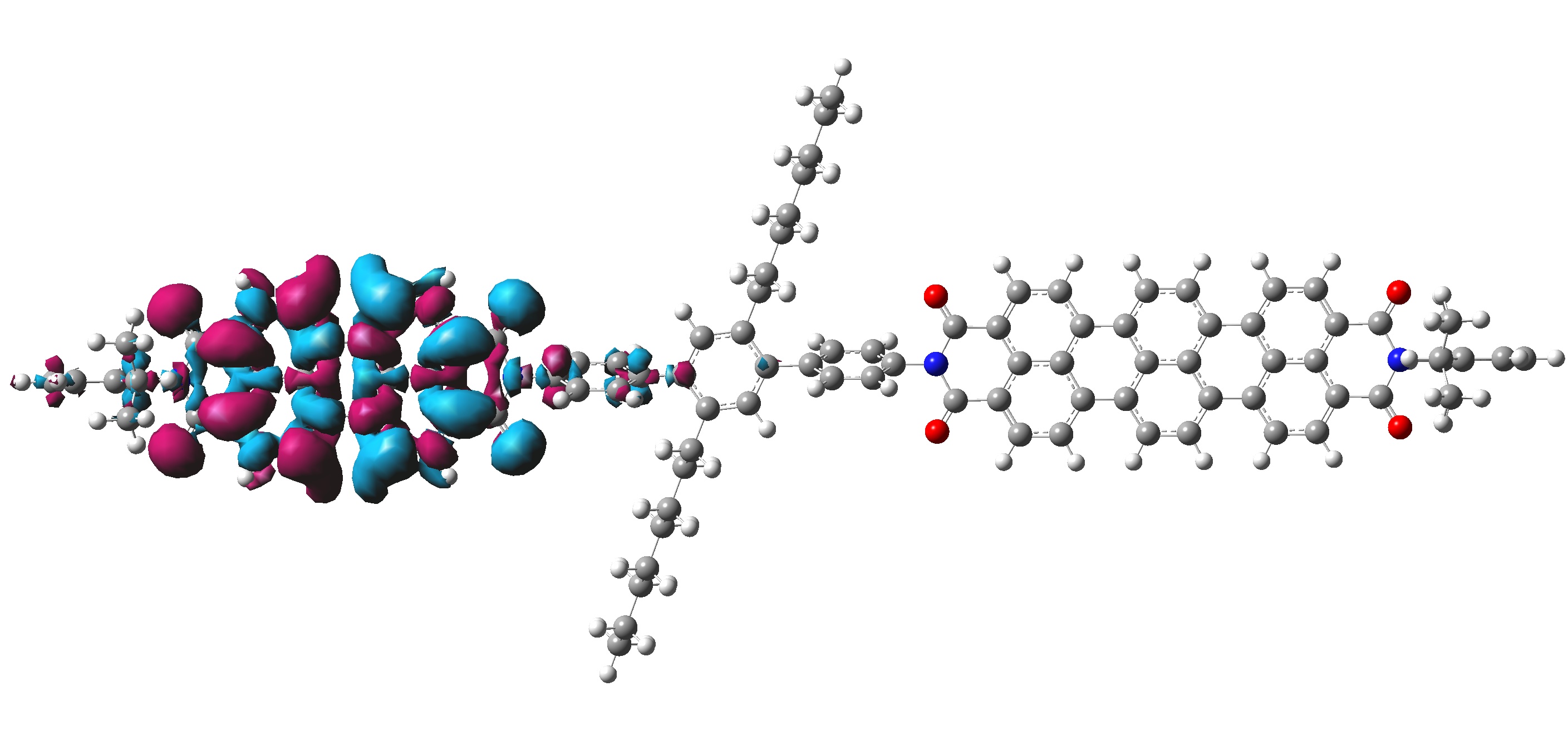} &
\includegraphics[scale=0.038]{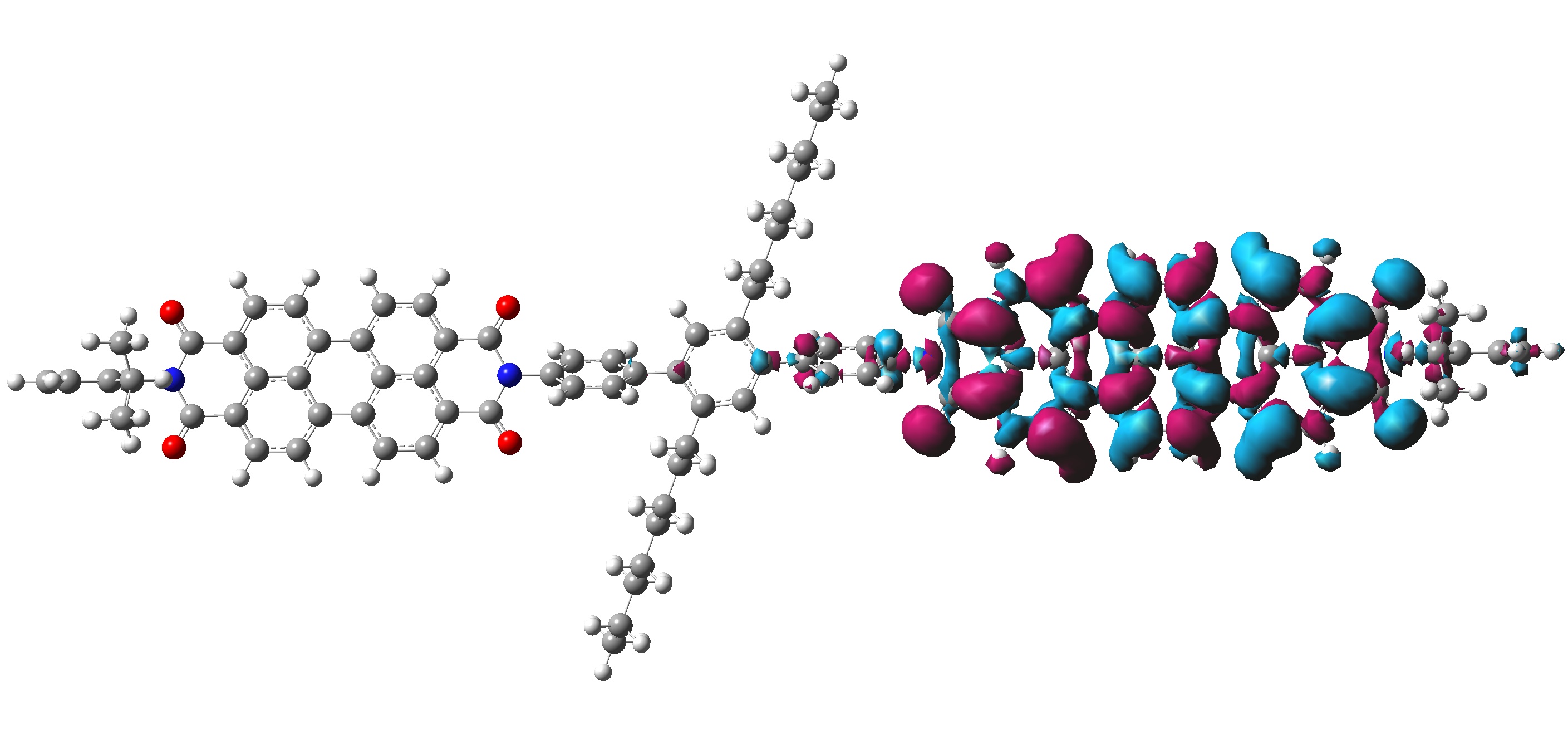}  \\
(c) Dyad 1 (${\rm S_0\rightarrow S_6}$)   & (d) Dyad 1 ( ${\rm S_0\rightarrow S_2}$)         \\
\hline
\includegraphics[scale=0.038]{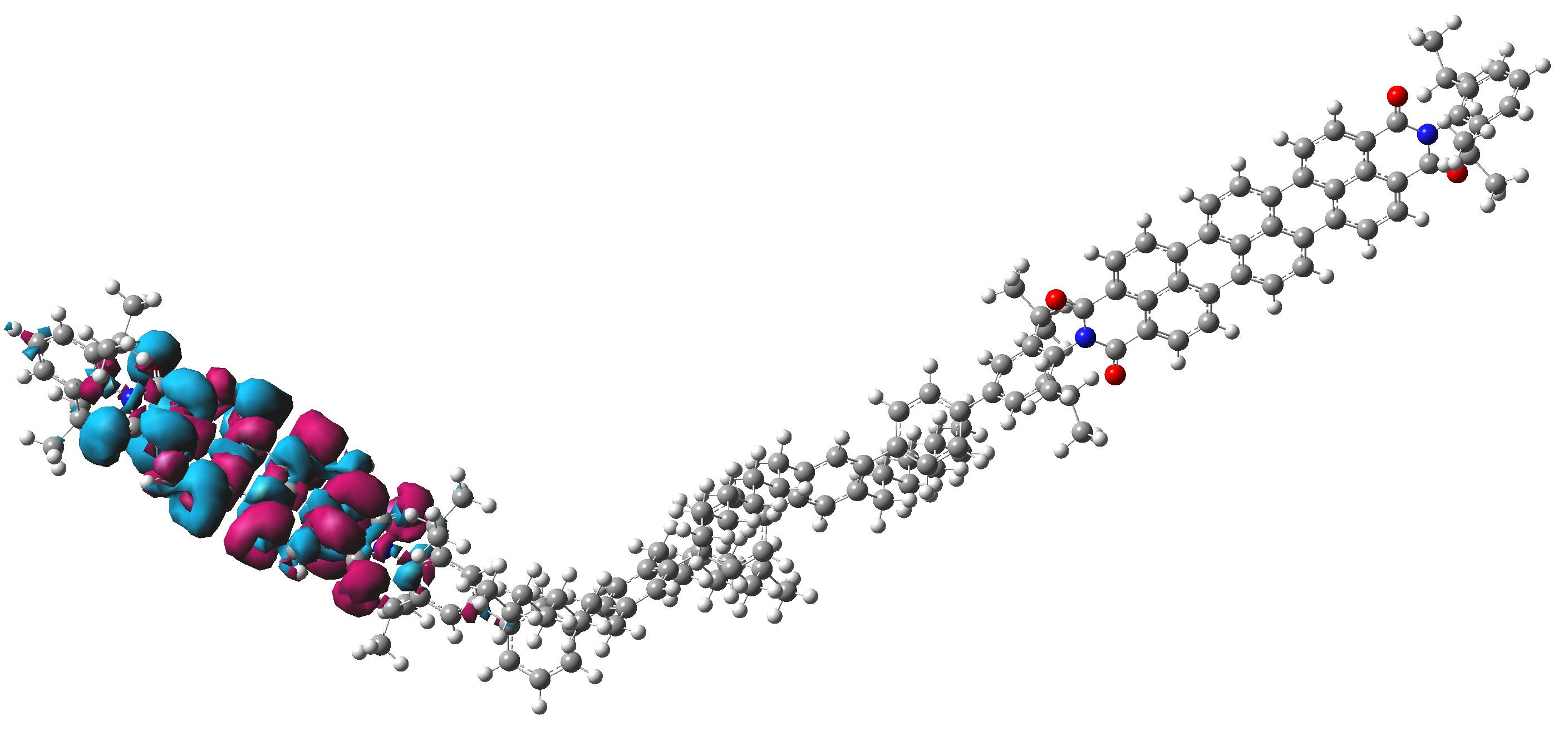} & 
\includegraphics[scale=0.038]{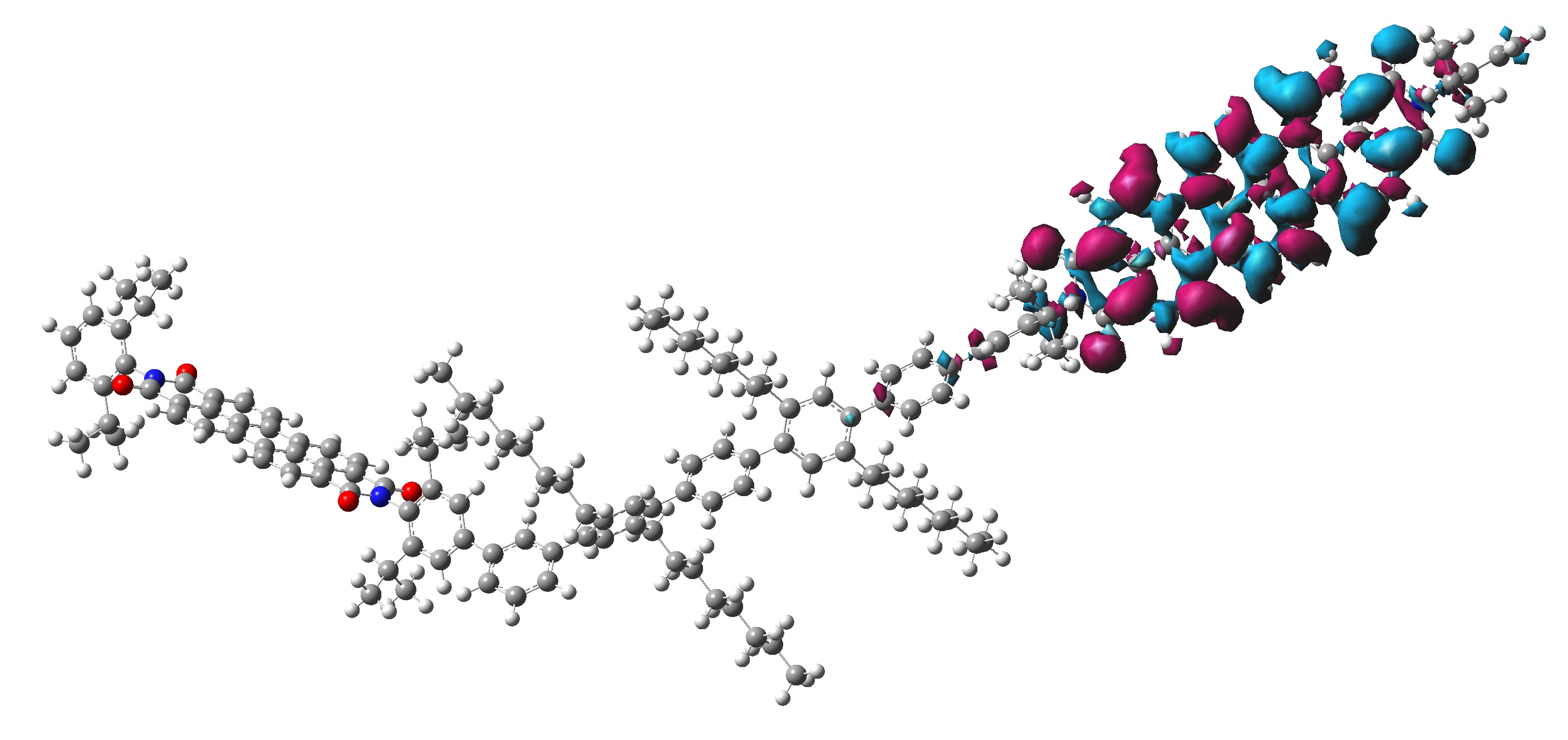}  \\
(e) Dyad 2 (${\rm S_0\rightarrow S_8}$) & (f) Dyad 2 (${\rm S_0\rightarrow S_2}$) \\
\hline
\end{tabular}
\caption{Transition densities for the electronic transitions denoted for structures optimized for the initial state.  The TD-DFT method with B3LYP functional was used.  }

\label{Transition-density}

\end{figure}

In order to examine the contribution to the exciton transfer of the bridge part, we conducted a series of calculations for different models for the donor and the acceptor.  These models include different numbers of  phenyl units and are denoted as  Ph-PDI-$n$Ph and $n$Ph-TDI-Ph ($n=1,\cdots,7$).
Note that there are octyl chains attached to the bridge backbones of the actual dyads used in experiments so as to enhance the solubility of the compound.
Assuming that the octyl chains have very little effect on the RET process, we replaced the bridge atoms and the octyl units with hydrogen atoms in these calculations.  
These added hydrogen atom bonds were optimized again before the excitation energies were calculated.  We also examined frontier molecular orbitals for each case. 
\re{The results (see Appendix A) show} that the highest occupied  and lowest unoccupied molecular orbitals remain almost the same visually for all of the corresponding parts, and that the frontier molecular orbitals are localized mostly at either PDI or TDI part.

The data in Table~\ref{Excitation-energy-nphenyl} show that the differences in excitation energies for different models of donor/acceptor are rather small.  However, an important trend can still be identified.  The excitation energies decrease with increasing $n$ as expected, approaching the values of the dyads. For example, the excitation energies of Ph-PDI-$n$Ph become red-shifted from $2.4206\ {\rm eV}$ to $2.4161\ {\rm eV}$ as $n$ increases form $1$ to $7$, approaching the corresponding excitation energy of {\bf D2}.  Similar trends can be found for $n$Ph-TDI-Ph, where red-shift of about $0.0046\ {\rm eV}$ occurs.  Thus, while the contributions of the bridging phenyl units to the excitation energies are very small, they are not negligible.  It is also important to note that the transition energies become much more insensitive to the length after $n=3\sim 4$.  The transition dipoles also show similar trends.   These suggest that those with $n=3\sim 4$ are reasonable choices as effective donor and acceptor units.

\begin{table}
\caption{\label{Excitation-energy-nphenyl} Calculated excitation energies (in ${\rm eV}$) for Ph-PDI-$n$Ph and $n$Ph-TDI-Ph ($n=1,\cdots, 7$) employing TD-DFT with B3LYP functional and 6-31G(d) basis. The values of oscillator strength (f) and the transition dipole moment in Debye (D) are also shown.}

       \begin{tabular}{c|c|ccc|ccc}

\hline
\hline
\multicolumn{2}{c|}{Molecule}               &\multicolumn{3}{c|}{\makebox[1.in]{ in Dyad 1}}  &\multicolumn{3}{c}{\makebox[1.in]{in Dyad 2}}    \\
\hline
                      &$n$-Ph &\makebox[.3in]{${\rm eV}$}  &\makebox[.3in]{$f$}    & \makebox[.3in]{D}&\makebox[.3in]{${\rm eV}$}  &\makebox[.3in]{$f$} &\makebox[.3in]{D}\\
        \hline
\multirow{8}*{Ph-PDI-}  &0 &2.4282     &0.6400 &\ 8.337 &2.4278    &0.6417 &\ 8.349  \\
  & 1 &2.4195    &0.7905 &\ 9.282 &2.4206    &0.7948 &\ 9.305  \\
   & 2 &2.4168    &0.8459 &\ 9.607&2.4174    &0.8587 &\ 9.678 \\
      & 3 &2.4161    &0.8728&\ 9.760 &2.4167    &0.8776  &\ 9.786  \\
   & 4 &&&&2.4163    &0.8880   &\ 9.843\\
   & 5 &&&&2.4162    &0.8947    &\ 9.882\\
    &6 &&&&2.4162    &0.8979    &\ 9.899\\
    &7 &&&&2.4161    &0.8989    &\ 9.905 \\
       \hline 
\multirow{8}*{-TDI-Ph}   & 0 &1.9667    &1.2054 &12.713 &1.9667    &1.2060  & 12.716\\
  & 1 &1.9560    &1.3160 &13.320  &1.9567    &1.3117   &13.295 \\
    &2 &1.9533    &1.3776&13.637 &1.9534    &1.3845  & 13.671 \\
    &3 &1.9525    &1.4123&13.811 &1.9525    &1.4233  & 13.864 \\
    &4 &&&&1.9522    &1.4438    &  13.965    \\
    &5 &&&&1.9522    &1.4561     &  14.024      \\
    &6 &&&&1.9521    &1.4628      &  14.057    \\
    &7 &&&&1.9521    &1.4655       &  14.070    \\
        \hline
        \hline
        \end{tabular}
\end{table}

\subsection{Electronic Couplings}
Based on the optimized structures as described in the previous subsection, we calculated electronic couplings using both the transition density cube (TDC) method\cite{krueger-jpcb102} and the following transition dipole (dp) approximation:  \begin{equation}
    J_{dp}=\frac{\kappa |\mbox{\boldmath $\mu$}_{D}| |\mbox{\boldmath $\mu$}_{A}|}{R^{3}}\ . \label{eq:j-dp}
\end{equation}
In the above expression, $\mbox{\boldmath$\mu$}_{D}$ and $\mbox{\boldmath$\mu$}_{A}$ are transition dipole vectors of the donor and the acceptor, respectively, $R$ is the distance between them, and $\kappa$ is the orientational factor defined as
\begin{equation}
      \kappa=\mbox{\boldmath$\hat \mu$}_{D} \cdot\mbox{\boldmath$\hat \mu$}_{A}-3(\mbox{\boldmath$\hat \mu$}_{D}\cdot {\bf \hat R})(\mbox{\boldmath$\hat \mu$}_{A}\cdot{\bf \hat R}) \ ,
\end{equation}
where $\mbox{\boldmath$\hat \mu$}_{D}$, $\mbox{\boldmath$\hat \mu$}_{A}$ and $\mbox{\boldmath$\hat R$}$ are unit transition dipole and distance vectors.

   As is well known, the dipole approximation becomes inaccurate when the size of the transition dipole moment of each molecule is comparable to the molecular separation. The TDC method, which accounts for the shape and the size of the transition density, produce more accurate values because it includes all the contributions of higher multipole transition moments by direct numerical integration of the product of electronic transition densities, although in an approximate way.   Let us label each cube element with volume $\delta x\delta y\delta z$ by $j$,  and denote the transition density for transition $\alpha$  of such cube element as $\rho_\alpha (x_j,y_j,z_j)$, where $x_j$, $y_j$, and $z_j$ are coordinates of the $j$th cube and  $\delta x$, $\delta y$, and $\delta z$ are lengths of the cube along respective Cartesian directions.   Then, the contribution of that element to the electronic transition is given by
\begin{equation}
M_{\alpha}^{j} = \rho_\alpha(x_j,y_j,z_j) \delta x \delta y \delta z \ .
\end{equation}
 The detailed expression for the transition density is as follows:
\begin{equation}
 \rho_{\alpha}(x,y,z)=\Psi_{\alpha g}(x,y,z)\Psi_{\alpha e}^{*}(x,y,z)  \ ,                
\end{equation}
where $\Psi_{\alpha g}$ and $\Psi_{\alpha e}$ represent the wave functions of the ground and excited electronic states, respectively.
Then, the Coulomb coupling between the two transitions $\alpha$ and $\alpha'$ within the TDC method is obtained by summing all the products of transition elements as follows:
\begin{equation}
J_{TDC}\cong\sum_{jk}\frac{M_{\alpha}^jM_{\alpha'}^k}{ r_{jk}}\ ,
\end{equation}
 where the summation is over all volume elements with major transitions densities for transitions $\alpha$ and $\alpha'$ and $r_{jk}$ is the separation between volume elements.

 Comparison of computational results between monomers and dyads, as shown in Table \ref{Excitation-energy-nphenyl}, confirms that Ph-PDI and TDI-Ph are the main contributors of the donor and acceptor, respectively, in both dyads. Figure~\ref{Transition-density} shows transition densities of both dyads and their monomers.
It is clear that the transition densities of ${\rm S_{0}\rightarrow S_{1}}$ of Ph-PDI and TDI-Ph constitute most of the transition densities of ${\rm S_{0}\rightarrow S_{6}}$, ${\rm S_{0}\rightarrow S_{2}}$ in {\bf D1}, and  most of those for ${\rm S_{0}\rightarrow S_{8}}$ and ${\rm S_{0}\rightarrow S_{2}}$ transitions in {\bf D2}, respectively.
The directions of the transition dipoles in both dyads are the same as those in separate monomers, which are both along the long axises of molecules.
However, it is worth pointing out that there are small portions of transition densities in both dyads located at intermediate phenyl bridges, which indicates that bridge units induce delocalization of the transition densities.  These contributions of bridge units, \re{although appear to be small}, cause significant change in electronic coupling.\cite{curutchet-jpcb112}

The electronic couplings between excited state of Ph-PDI and the ground state of TDI-Ph are listed in Table~\ref{couplings}, together with the distance between centers of transition densities $R$ and the orientational factor $\kappa$.  For {\bf D2},  the magnitude of the electronic coupling calculated by the TDC method is relatively close to that calculated by the transition dipole approximation, which means that the finite size effect of transition densities is rather small.  However,  for {\bf D1}, the discrepancy between the two shows that there are significant contributions of higher order multipolar components.  The data in Table~\ref{couplings} show that the non-dipole effect in this case results in about 20 \% increase of the electronic coupling.  

Actual electronic couplings in toluene or PMMA are expected to be different from those in Table~\ref{couplings} due to medium effects. In a simple manner, this may be accounted for by incorporating the bulk solvation effect \re{through refractive index}.  Either employing the refractive index of PMMA, $n_r=1.491$,  or that of toluene $n_r=1.496$, the screening factor due to the bulk solvent effect is $s=1/n_r^{2}=0.45$.   This is slightly smaller than the value of $0.487$ obtained from more advanced models of solvation.\cite{caprasecca-jctc8}  The electronic couplings between the donor and acceptor calculated by the TDC method for {\bf D1} and {\bf D2}, when multiplied by this screening factor, respectively become $-28.0\ {\rm cm^{-1}}$ and $-6.82\ {\rm cm^{-1}}$.   Similar level of reduction is expected for all other electronic coupling constants.

\begin{table}
\caption{\label{couplings} Electronic coupling constants  in ${\rm cm^{-1}}$ calculated by the TDC method. The values in parentheses are those based on transition dipole approximation. Here, the donor (D) represents Ph-PDI, the acceptor (A) represents TDI-Ph, and the bridge ($B_k$) with $k=1,2$ respectively represents the first and the second excitations of the oligophenylene units of the two dyads.   Distances between D and A and orientational factors are also shown.}

     \begin{tabular}{cccc|c|ccc}

\hline
\hline
\makebox[.5in]{Molecule}  &  \makebox[.3in]{${\rm R}$(${\rm \AA}$) }  &\makebox[.2in]{$\kappa^{2}$}   &\makebox[.5in]{$J_{DA}$} &\makebox[.2in]{$k$}&\makebox[.4in]{$J_{DB_k}$} & \makebox[.5in]{$J_{B_kA}$}  \\
        \hline
\multirow{2}*{Dyad 1} & \multirow{2}*{29.33}     &\multirow{2}*{4.0}            &\multirow{2}*{-57.5 (-45.9)}&1 &  -271 & -549 \\
                                   &                                      &                                        &                                          &2 & -261 &  -90 \\
                                   \hline
\multirow{2}*{Dyad 2} & \multirow{2}*{41.43}     &\multirow{2}*{2.5}            &\multirow{2}*{-14.0  (-12.9)}&1 &  -140 & 574  \\
                                   &                                      &                                        &                                           &2 & 38      & 119 \\ 
        \hline
        \hline
        \end{tabular}
\end{table}

\begin{table} 
\caption{\label{Bridge-ex} Calculated excitation energies (in ${\rm eV}$) for the bridge unit of {\bf D1}  employing TD-DFT with three different functionals and 6-31G(d) basis.  Computational data for ${\rm S_0\rightarrow S_1}$ (or ${\rm S_2}$) were calculated for optimized structures for the ground electronic state (${\rm S_0}$), and those for the ${\rm S_1\rightarrow S_0}$ were calculated for optimized structures for the first excited singlet state (${\rm S_1}$).}

\begin{tabular}{c|c|l|c|c}
\hline
\hline
Transitions & Method      &env.         &\makebox[.6in]{$f$} &\makebox[.6in]{${\rm E_{cal} \ (eV)}$}              \\
        \hline
\multirow{6}*{${\rm S_{0}\rightarrow S_{1}}$} &\multirow{2}*{B3LYP}    &v.   &   0.2194 & 4.7755                  \\

                                                               &                                      &s. &0.4084 &4.7311                       \\
                                                               \cline{2-5}
                                                               &\multirow{2}*{CAM-B3LYP}    &v.   &  0.0621  &     5.2032             \\

                                                               &                                      &s. &0.1530 & 5.1834                      \\
                                                               \cline{2-5}
                                                               &\multirow{2}*{M06-2X}    &v.   &  0.1358  & 5.1695                 \\

                                                               &                                      &s. & 0.3291& 5.1374                      \\

\hline
\multirow{6}*{${\rm S_{0}\rightarrow S_{2}}$}     &\multirow{2}*{B3LYP}    &v.   & 0.2002   & 4.9375                   \\

                                                               &                                      &s. &0.1967 & 4.9097                       \\
                                                               \cline{2-5}    
                                                                &\multirow{2}*{CAM-B3LYP}      &v.  &0.4027 &5.4440\\     
          
         &      &s. &  0.5553  & 5.3835                               \\ 
         
\cline{2-5}         
           &\multirow{2}*{M06-2X}          &v.   & 0.4316 &5.3662      \\

  & &s.& 0.4844 & 5.3138 \\
       \hline      
 \multirow{7}*{${\rm S_{1}\rightarrow S_{0}}$}&\multirow{2}*{B3LYP}    &v.   &0.7125&3.6051   \\

  & &s. &0.9890 &3.4511                        \\
 \cline{2-5} 
 &B3LYP/CIS & v. & 0.7742 & 3.6856     \\
 \cline{2-5}
            &\multirow{2}*{CAM-B3LYP} &v.&0.7679&3.8775                                  \\
         
          &&s. & 1.0450& 3.7254  \\ 
\cline{2-5}
                    &\multirow{2}*{M06-2X}  &v.   &0.7674&3.8853                        \\
                   
                   &&s. &1.0538 &3.7375       \\
                   \hline
       
   & \multirow{2}*{B3LYP}  &       v. &    & 4.1082                          \\
 
 &&s. &  &3.9857  \\
 \cline{2-5}
$0-0$       &\multirow{2}*{CAM-B3LYP} &v.& &4.5148   \\
  
 $({\rm S_0-S_1})$ &&s. & & 4.3901  \\
  \cline{2-5}
   &\multirow{2}*{M06-2X} &v. & & 4.4843 \\
   
   &&s.& & 4.3612 \\
       \hline
       \hline
         \end{tabular}
\end{table}

Another key issue in determining the donor-acceptor electronic coupling is to understand the contribution of the bridge unit.  Table \ref{couplings} also provides Coulomb couplings between the bridge units (modeled as separate entities) and Ph-PDI or TDI-Ph, respectively, for each case of {\bf D1} and {\bf D2}.   These do not include exchange and overlap interactions, but are yet an order of magnitudes larger than those between Ph-PDI and TDI-Ph parts of dyads.  The net contribution of these to the effective donor-acceptor electronic coupling can be accounted for approximately by using the following expression normally known as super-exchange interaction:\cite{chen-jcp129} 
\be
J_{se}=\sum_k \frac{J_{DB_k}J_{B_kA}}{E_D-E_{B_k}} \ ,  \label{eq:j_se}
\ee
which in fact can be derived from the second order time dependent perturbation theory.  In the above expression, $B_{k}$ represents the bridge unit in the $k$th excited state and $E_{B_k}$ the corresponding excitation energy.  For the evaluation of Eq. (\ref{eq:j_se}),  it is necessary to calculate the energies of these excited states  of the bridge unit.  Table  \ref{Bridge-ex} provides the absorption, emission, and 0-0 transition energies for the bridge unit of {\bf D1}.  

Assuming that the relative energy of the bridge unit is equal to the difference between the absorption energies calculated employing the B3LYP functional in vacuum, the resulting super-exchange couplings due to the first and the second excited states of the bridge unit in {\bf D1} are respectively, $-7.86$ and $-1.13\ {\rm cm^{-1}}$.    If differences in emission energies are used instead, these values become larger.  For example, the contribution from the first excited state of the bridge becomes  $-13.45\ {\rm cm^{-1}}$.  These are already significant, and further inclusion of the contributions of higher excited states of the bridge unit can result in substantial change in the effective donor-acceptor electronic coupling.  However, considering the approximations involved in Eq. (\ref{eq:j_se}) and the \re{large reorganization energy of the bridge unit, as can be seen from the difference between their absorption and emission transition energies in Table \ref{Bridge-ex}, it is not clear whether the estimates calculated this way are accurate enough.}  

\begin{table}
\caption{\label{Coupling-ab} Magnitudes of electronic couplings in vacuum calculated by the TDC method and dipole approximation (in parentheses) for Ph-PDI-$n_D$Ph and $n_A$Ph-TDI-Ph in ${\rm cm^{-1}}$ for Dyad 1  (I) and Dyad 2 (II). }

       \begin{tabular}{c|c|cccccccc}

\hline
\hline
&&\multicolumn{8}{c}{$n_A$}\\
\cline{2-10}
&$n_D$&0        &1       &2       &3    &4 &5&6&7  \\
\hline
\multirow{8}*{I}&\multirow{2}*{0}        &53.0&61.1&71.4&89.0\\
  &      &(42.3)&(46.4)&(48.7)&(50.4)\\
        \cline{2-6}
&\multirow{2}*{1}        &64.9&76.1&92.4&         \\
  &    &(48.9)&(53.8)&(56.4)&         \\
      \cline{2-5}
&\multirow{2}*{2}        &76.6&93.2&          &         \\
  &     &(51.8)&(57.0)&          &         \\
       \cline{2-4}
&\multirow{2}*{3}        &94.1&          &          &         \\
  &     &(53.5)&          &          &         \\
\hline
\multirow{16}*{II}&\multirow{2}*{0}      &12.9&14.1&15.3&16.3&17.2&18.1&18.7&20.6\\
  &      &(11.9)&(12.8)&(13.5)&(13.8)&(14.1)&(14.2)&(14.3)&(14.4)\\
        \cline{2-10}
&\multirow{2}*{1}        &14.6&16.0&17.4&18.6&19.7&20.9&22.0&       \\
  &      &(13.4)&(14.5)&(15.2)&(15.7)&(15.9)&(16.1)&(16.2)&       \\
        \cline{2-9}
&\multirow{2}*{2}        &15.3&16.8&18.3&19.6&20.9&22.6&          &       \\
  &     &(14.1)&(15.2)&(16.0)&(16.5)&(16.7)&(16.9)&          &       \\
       \cline{2-8}
&\multirow{2}*{3}        &16.1&17.7&19.3&20.8&22.6&          &          &       \\
  &     &(14.5)&(15.6)&(16.4)&(16.9)&(17.1)&          &          &       \\
       \cline{2-7}
&\multirow{2}*{4}        &17.0&18.7&20.6&22.7&          &          &          &       \\
  &     &(14.7)&(15.9)&(16.7)&(17.2)&          &          &          &       \\
       \cline{2-6}
&\multirow{2}*{5}        &18.0&20.0&22.6&          &          &          &          &       \\
  &      &(14.9)&(16.1)&(16.9)&          &          &          &          &       \\
        \cline{2-5}
&\multirow{2}*{6}        &19.5&22.2&          &          &          &          &          &       \\
  &      &(15.0)&(16.2)&          &          &          &          &          &       \\
        \cline{2-4}
&\multirow{2}*{7}        &22.5&          &          &          &          &          &          &       \\
  &      &(15.1)&          &          &          &          &          &          &       \\
\hline
\hline
        \end{tabular}
\end{table}

Alternatively, one can include parts of the bridge units as extended donor and acceptor along with Ph-PDI and TDI-Ph by extending the approach taken by Curutchet {\it et al.}.\cite{curutchet-jpcb112}  We have already provided the excitation energies and transition dipoles of such extended units of varying length in Table \ref{Excitation-energy-nphenyl}.   As expected, the transition dipole moment increases as the number of phenyl units $n$ increases. For example, in models constituting {\bf D1}, as $n$ increases from 0 to 3, the transition dipole moments increase from 8.337 to 9.760 Debye for Ph-PDI-$n$Ph, and from 12.731 to 13.811 Debye  for $n$Ph-TDI-Ph. In models constituting {\bf D2}, as $n$ increases from 0 to 7, they change from 8.349 to 9.905 Debye for Ph-PDI-$n$Ph, and from 12.716 to 14.070 Debye for $n$Ph-TDI-Ph .  These increases indicate that the state of the bridge units participate in electronic transitions corresponding to effective donor and acceptor. Interestingly, in  {\bf D2}, the value of the transition dipole \re{starts becoming insensitive to $n$ before it reaches $7$}, which suggests that \re{parts of the bridge unit smaller than its actual length in {\bf D2} can be used to define effective sizes of  donor or acceptor excitons}.   Based on the same structural data and electronic structure calculation results,  we have also calculated corresponding Coulomb interactions between the series of \re{effective} donors and acceptors by both the TDC method and transition dipole approximation. \re{The results} are listed in Table~\ref{Coupling-ab}.   

Although the distances between Ph-PDI and TDI-Ph parts remain the same and most of transitions are localized in these parts, it is important to note that the couplings between the extended donor and acceptor units increase  as $n_D+n_A$ becomes larger.  This is consistent with the fact that the transition dipoles increase as \re{more bridge units are included}.  Comparison of the TDC results with those based on the dipole approximation also shows that higher order multipolar components \re{of those increased donor and acceptor units} make additional positive contributions \re{to effective donor-acceptor electronic couplings}.  

Another interesting and \re{quite} revealing trends are that the effective electronic couplings vary little within the series of fixed value of $n_D+n_A$, as \re{has been observed in a previous study}\cite{curutchet-jpcb112} for {\bf D1}.   As yet, we find that the maximum value of electronic coupling is obtained for $n_D=3$ in {\bf D1} and for $n_D=4$ in {\bf D2}.  These values correspond to those where the values of excitation energy and transition dipole moment saturate and do not change significantly with further increase of the bridge unit.  Therefore, we \re{conclude} that Ph-PDI-3Ph can be best viewed as an effective donor in {\bf D1} and Ph-PDI-4Ph as an effective donor in {\bf D2}.    \vspace{.1in}\\

\re{Based on} the results of calculation reported so far, \re{which include Coulomb interaction contributions of the bridge unit,} we can estimate that the effective donor-acceptor \re{electronic} coupling for {\bf D1} is a factor of 2.05 larger than \re{that based on }the transition dipole interaction between the Ph-PDI and TDI-Ph unit.  For the case of {\bf D2}, the enhancement factor is 1.76.   The medium effects are not expected to change these factors significantly considering that the screening factor calculated by full polarizable model and that using simple bulk refractive index are similar.    Thus, if we assume that experimentally observed non-FRET behavior is purely due to the enhancement of the electronic coupling, these imply enhancement of the FRET rate by  about a factor of $4.2$ for {\bf D1} and by about a factor of $3.2$ for {\bf D2}.    While \re{such enhancement} is sufficient to explain experimental results for {\bf D2}, it falls short of explaining the experimental results for {\bf D1} by about a factor of 2.   This suggests \re{it is necessary to consider other mechanisms of non-FRET behavior}. 

\section{non-FRET rate mechanisms}
\noindent
For {\bf D1}, a number of different non-FRET mechanisms can be explored. While we have developed\cite{jang-jcp129,jang-jcp131,jang-jcp135,yang-jcp137,jang-exciton} and applied\cite{yang-jacs132,jang-exciton,jang-wires3,jang-njp15,jang-rmp90} theories of coherent RET by combining a second order quantum master equation approach with polaron transformation,\cite{holstein-ap8-1,holstein-ap8-2, rackovsky-mp25} \re{we do not expect application of such theories is necessary for the dyads considered here because the large discrepancy between the donor and acceptor excitation energies compared to other parameters is likely to make the effect of quantum coherence very small.}  There is no experimental evidence supporting significant contribution of quantum coherence either.   Thus, consideration within the rate behavior seems sufficient \re{for this case}.    

There are three possible mechanisms of non-FRET rate processes warranting detailed examination.  First, there can be significant effects of common modes.\cite{jang-cp275,hennebicq-jcp130}  Second, inelastic RET process,\cite{jang-jcp127,yang-jcp137} namely, quantum modulation of electronic couplings via environmental degrees of freedom, can also enhance the rate. Third, contributions of dark exciton states are possible, which can be accounted for through the multichromophoric FRET (MC-FRET) theory\cite{jang-prl92} or its non-Markovian generalization.\cite{jang-prl113} Considering the downhill energetic condition, this is plausible if there are dark higher exciton states of the TDI-Ph part with significant electronic couplings to the $S_1\rightarrow S_0$ transition of the (extended) donor.  We have examined this possibility by calculating the Coulomb interactions between the $S_1\rightarrow S_0$ transition of Ph-PDI-3Ph and $S_0\rightarrow S_k\ (k\geq 2)$ transitions of the TDI-Ph unit.   The values we obtained are $2.617$, $0.029$,  $4.907$, and $0.0127\ {\rm cm^{-1}}$ for $k=2-5$.   Considering the smallness of these values, we concluded that the contributions of higher dark excited states  of TDI-Ph can be ignored.  

The effects of common modes and inelastic process can be addressed in terms of an effective two state model.  Considering the fact that the bridge unit of the dyad is included as parts of the donor and the acceptor in this model, there can be significant number of vibrational modes, which  are coupled to both the donor and the acceptor and cannot entirely be decoupled into those constituting emission and absorption lineshapes. In addition, the flexibility of the bridge unit, which can modulate the magnitude of electronic coupling, \re{may lead to significant inelastic RET process}.  \re{In order to consider these two effects}, let us assume the following effective two-state model:
 \begin{eqnarray}
\hat H&=&E_D|D\rangle\langle D|+E_A|A\rangle\langle A|+\hat J(|D\rangle\langle A|+|A\rangle\langle D|)  \nonumber \\
&+&\hat X_D|D\rangle\langle D|+\hat X_A|A\rangle\langle A| +\hat H_b \ ,\label{eq:hamil-inelastic}
\end{eqnarray}
where $E_D$ and $E_A$ are vertical excitation energies of $D$ and $A$, $\hat J$ is the electronic coupling between $|D\rangle$ and $| A\rangle$, $\hat X_D$ and $\hat X_A$ are bath operators coupled  to $|D\rangle$ and $|A\rangle$, and $\hat H_b$ is the bath Hamiltonian.  Note that the bath here includes both molecular vibrations and all the environmental degrees of freedom and that operator notation $\hat J$  reflects that it in general depends on the bath degrees of freedom.     
The Fermi Golden Rule (FGR) rate for the transfer of exciton  for the above Hamiltonian can be expressed as\cite{jang-exciton}
\ben
&&k_{FG}=\frac{2}{\hbar^2} {\rm Re} \left [ \int_0^\infty dt\ e^{i(E_D-E_A)t/\hbar} \right . \nonumber \\ 
&&\left . \times Tr_b\left \{ e^{i(\hat H_b+\hat X_D)t/\hbar} \hat J^\dagger e^{-i(\hat H_b+\hat X_A)t/\hbar} \hat J \hat \rho_{b,s}^D\right \}\right ]  \ , \label{eq:kfg-1}
\een
where
\be
\hat \rho_{b,s}^D=\frac{e^{-\beta (\hat H_b+\hat X_D)}}{Tr_b \left \{ e^{-\beta (\hat H_b+\hat X_D)} \right \}}\ .
\ee

Let us assume that the bath Hamiltonian can be represented by a set of harmonic oscillators and that $\hat X_D$ and $\hat X_A$ are linear in the displacements of the oscillators as follows: 
\begin{eqnarray}
&&\hat H_b=\sum_{n} \hbar\omega_{n} (\hat b_{n}^\dagger \hat b_{n}+\frac{1}{2}) \ ,  \label{eq:hb-har}\\
&&\hat X_D=\sum_{n} \hbar\omega_{n} (\hat b_{n}+\hat b_{n}^\dagger)g_{nD} \ , \label{eq:xd-har} \\
&&\hat X_A= \sum_{n} \hbar\omega_{n} (\hat b_{n}+\hat b_{n}^\dagger)g_{nA}  \ ,  \label{eq:xa-har}       
\end{eqnarray}
where $\hat b_{n}$ and $\hat b_{n}^\dagger$ are lowering and raising operators of the $n$th normal modes of the bath, and $g_{nD}$ and $g_{nA}$ represent dimensionless coupling strengths  of the $n$th mode to $|D\rangle$ and $|A\rangle$, respectively.
The net effect of these couplings can be specified by the following three spectral densities:
\ben
&&{\mathcal J}_D(\omega)=\pi\hbar \sum_n \delta (\omega-\omega_n)\omega_n^2 g_{nD}^2 \ , \\
&&{\mathcal J}_A(\omega)=\pi\hbar \sum_n \delta (\omega-\omega_n)\omega_n^2 g_{nD}^2 \ , \\
&&{\mathcal J}_c(\omega)=-\pi\hbar \sum_n \delta (\omega-\omega_n)\omega_n^2 g_{nD}g_{nA} \ , 
\een
where ${\mathcal J}_D(\omega)$ and ${\mathcal J}_A(\omega)$ are bath spectral densities of $D$ and $A$ respectively, and ${\mathcal J}_c(\omega)$ represents the contribution of common modes.   While ${\mathcal J}_D(\omega)$ and ${\mathcal J}_A(\omega)$ are nonnegative, ${\mathcal J}_c(\omega)$ can take both positive and negative values under the condition that $|{\mathcal J}_c(\omega)| \leq {\mathcal J}_D(\omega)+{\mathcal J}_A (\omega)$.   The final bath spectral density that enters the FGR rate expression is the following spectral density: 
\ben
{\mathcal J}_s(\omega)&=&\pi\hbar \sum_n \delta (\omega-\omega_n)\omega_n^2 (g_{nD}-g_{nA})^2 \nonumber \\
&=&{\mathcal J}_D(\omega)+{\mathcal J}_A(\omega)+2{\mathcal J}_c(\omega) \ . 
\een
Thus, positive ${\mathcal J}_c(\omega)$ amounts to net increase of ${\mathcal J}_s(\omega)$.  

For each of the spectral density, let us also define the reorganization energy, and real and imaginary parts of the lineshape function as follows:
\ben
&&\lambda_\gamma=\frac{1}{\pi}\int_0^\infty d\omega \frac{{\mathcal J}_\gamma(\omega)}{\omega}\ , \\
&&G_{R,\gamma} (t)=\frac{1}{\pi\hbar}\int_0^\infty d\omega \frac{ {\mathcal J}_{\gamma}(\omega)}{\omega^2}\coth \left (\frac{\beta\hbar\omega}{2}\right )(1-\cos (\omega t) )\ ,\nonumber \\ \\
&&G_{I,\gamma} (t)=\frac{1}{\pi\hbar}\int_0^\infty d\omega \frac{ {\mathcal J}_{\gamma}(\omega)}{\omega^2} (\sin (\omega t) -\omega t)  \nonumber  \\
&&\hspace{.5in}=\frac{1}{\pi\hbar}\int_0^\infty d\omega \frac{ {\mathcal J}_{\gamma}(\omega)}{\omega^2} \sin (\omega t) -\frac{\lambda_\gamma}{\hbar} t\ ,
\een
where $\gamma=D$, $A$, $c$, or  $s$. 
 Then, it is straightforward\cite{jang-exciton} to show that the normalized absorption and emission lineshape functions (for unit transition dipole moment) are given by 
\ben
I_\alpha(\omega)&=&\frac{1}{2\pi}\int_{-\infty}^\infty dt\ \exp \Bigg\{i \left (\omega -\frac{E_\alpha}{\hbar} \right )t\nonumber \\
&&\hspace{.3in}-G_{R,\alpha}(t)-iG_{I,\alpha}(t) \Bigg\} , \\
{\mathcal E}_\alpha(\omega)&=&\frac{1}{2\pi}\int_{-\infty}^\infty dt\ \exp \Bigg\{i \left (\omega -\frac{(E_\alpha -2\lambda_\alpha)}{\hbar} \right )t\nonumber \\
&&\hspace{.3in}-G_{R,\alpha}(t)+iG_{I,\alpha}(t)\Bigg\}  , 
\een 
where $\alpha=D$ or $A$. For strongly coupled bath in the high temperature limit, 
\be
G_{R,\alpha}(t) +iG_{I,\alpha}(t) \approx \frac{\lambda_\alpha}{\beta\hbar^2}t^2 .
\ee 
Therefore, the above lineshapes in this limit reduce to the following Gaussian forms:
\ben
&&I_\alpha(\omega)\approx \left (\frac{\beta\hbar^2}{4\pi\lambda_\alpha}\right )^{1/2}\exp \Bigg \{-\frac{\beta\hbar^2}{4\lambda_\alpha} \left (\omega -\frac{E_\alpha}{\hbar}\right )^2 \Bigg \} ,\\
&&{\mathcal E}_\alpha(\omega)\approx \left (\frac{\beta\hbar^2}{4\pi\lambda_\alpha}\right )^{1/2}\exp \Bigg \{-\frac{\beta\hbar^2}{4\lambda_\alpha} \left (\omega -\frac{E_\alpha-2\lambda_\alpha}{\hbar}\right )^2 \Bigg \} . \nonumber\\
\een

In general, one may introduce the following models for the bath spectral densities: 
\ben 
&&{\mathcal J}_D(\omega)={\mathcal J}_{_{\rm Ph-PDI}} (\omega)+f_{_D} {\mathcal J}_B(\omega)+s_{_D} {\mathcal J}_{env}(\omega), \label{eq:jd}\\ 
&&{\mathcal J}_A(\omega)={\mathcal J}_{_{\rm TDI-Ph}} (\omega)+f_{_A} {\mathcal J}_B(\omega)+s_{_A} {\mathcal J}_{env}(\omega) ,  \label{eq:ja}\\
&&{\mathcal J}_c(\omega)=f_c(f_{_D}+f_{_A}) {\mathcal J}_B(\omega) +s_c(s_{_D}+s_{_A})   {\mathcal J}_{env}(\omega) ,\nonumber\\   \label{eq:jb}
\een
where  ${\mathcal J}_{_{\rm Ph-PDI}}(\omega)$, ${\mathcal J}_{_{\rm TDI-Ph}} (\omega)$, and ${\mathcal J}_B(\omega)$ are bath spectral densities coming from the molecular vibrations of Ph-PDI, TDI-Ph, and the bridge unit. ${\mathcal J}_{env}(\omega)$ represents all the environmental degrees of freedom.  Other parameters in above equations represent \re{extents of contributions of these components to the donor, acceptor, and common bath spectral densities},  and are assumed to satisfy the following conditions: $0\leq f_{_D}, f_{_A} ,s_{_D}, s_{_A}\leq 1$ and  $-1\leq f_c,s_c \leq 1$.  This latter condition ensures that ${\mathcal J}_c(\omega)$ remains bounded by ${\mathcal J}_D(\omega)+{\mathcal J}_A(\omega)$ at each frequency. 

\begin{figure}
\includegraphics[scale=0.45]{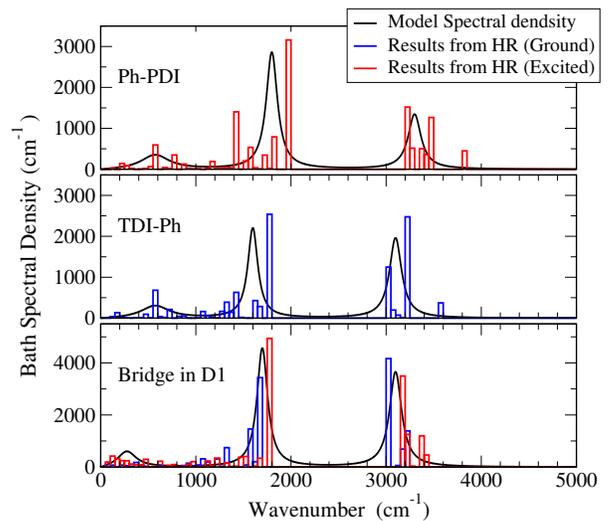} 
\caption{\label{spd-mod}Spectral densities modeling Ph-PDI, TDI-Ph, and the bridge unit of {\bf D1} with respect to those derived directly from a coarse grained sum (with an interval of $50\ {\rm cm^{-1}}$) of Huang-Rhys factors weighted by squares of corresponding wavenumbers.  Each model spectral density consists of three Brownian-oscillator spectral densities as shown in Eq. (\ref{eq:spec-bo}).  The ones obtained from Huang-Rhys factors were scaled by arbitrary factors to visually match the maximum intensities. }   
\end{figure}

\re{There are various approaches to model the bath spectral densities of Ph-PDI, TDI-Ph, and the bridge unit.\cite{jang-rmp90,tong-jcp153}} In this work, we have first calculated Huang-Rhys factors from the displacements between ground and excited states for all the vibrational modes\cite{reimers-jcp115} and then determined spectral densities that are coarse-grained over an interval of $50\ {\rm cm^{-1}}$. Figure 5 shows these spectral densities as histograms.  Then, we have approximated these as sums of three Brownian oscillator models as follows:
\be
{\mathcal J}_{BO}(\omega)=\pi \hbar\sum_{k=1}^3 \frac{\eta_k \Omega_k^4 \omega }{\left (\omega^2-\Omega_k^2\right )^2+4\omega^2\Gamma_k^2} \ . \label{eq:spec-bo}
\ee
The resulting \re{model} spectral densities are also shown in Fig. \ref{spd-mod}  and the values of parameters for these models are provided in Table \ref{table-bo}.  
Note that \re{some of the environmental effects are included in these Brownian oscillator models.} 

\begin{table}
\caption{\label{table-bo} Parameters for three sets of Brownian oscillators modeling the bath spectral densities derived from the Franck-Condon factors of Ph-PDI, TDI-Ph, and the bridge unit of {\bf D1}.  The reorganization energies for these are respectively assumed to be $781$, $675$, and $4720\ {\rm cm^{-1}}$ respectively.}
\begin{tabular}{c|c|c|c}
\hline
\hline
&\makebox[0.7in]{Ph-PDI}&\makebox[0.7in]{TDI-Ph}&\makebox[0.7in]{Bridge in {\bf D1}}\\
\hline
$\eta_1$&0.2086 &0.1805 & 1.2604\\
$\Omega_1 ({\rm cm^{-1}}) $&600 &600 & 300 \\
$\Gamma_1 ({\rm cm^{-1}})$& 210 & 210&150 \\
\hline
$\eta_2$&0.01252 &0.01083 & 0.01765\\
$\Omega_2 ({\rm cm^{-1}}) $&1800 &1600 & 1700 \\
$\Gamma_2 ({\rm cm^{-1}})$& 81 &72 &76.5 \\
\hline
$\eta_3$&0.001043 &0.001625 & 0.003025\\
$\Omega_3 ({\rm cm^{-1}}) $&3300 &3100 &3100  \\
$\Gamma_3 ({\rm cm^{-1}})$&82.5  &77.5 &77.5 \\
\hline
\hline
\end{tabular}
\end{table}

For the case where the electronic coupling is independent of the bath degrees of freedom and remains constant, Eq. (\ref{eq:kfg-1}) simplifies to 
\ben
&&k_{FG}=\frac{|J|^2}{\hbar^2} {\rm Re} \left [ \int_0^\infty dt\ e^{i(E_D-E_A)t/\hbar} \right . \nonumber \\ 
&&\left . \times Tr_b\left \{ e^{i(\hat H_b+\hat B_D)t/\hbar} e^{-i(\hat H_b+\hat B_A)t/\hbar} \hat \rho_{b,s}^D\right \}\right ]   . \label{eq:kfg-2}
\een
This expression can be simplified further by employing explicit expressions for the linearly coupled bath of harmonic oscillators, Eqs. (\ref{eq:hb-har})-(\ref{eq:xa-har}), and can be expressed as\cite{jang-exciton}  
\ben
k_{FG}&=&\frac{|J|^2}{\hbar^2} \int_{-\infty}^\infty dt \exp \Bigg \{\frac{i}{\hbar}(E_D-\lambda_D-E_A+\lambda_A) t \nonumber \\
&&\hspace{.5in}-G_{R,s}(t)-iG_{I,s}(t) -\frac{i}{\hbar}\lambda_s t\Bigg \} \  \nonumber \\
&=&\frac{|J|^2}{\hbar^2} \int_{-\infty}^\infty dt \exp \Bigg \{\frac{i}{\hbar}(E_D-2\lambda_D-E_A) t \nonumber \\
&&\hspace{.5in}-G_{R,s}(t)-iG_{I,s}(t) -\frac{i}{\hbar}\lambda_c t\Bigg \}  , \label{eq:k-fg-common}
\een 
where the fact that $\lambda_s=\lambda_D+\lambda_A+\lambda_c$ has been used in the second line.  For the case where $\lambda_c$ is positive, the above expression can be converted to the spectral overlap expression similar to F\"{o}rster's by employing the fact that $G_{R,s}(t)=G_{R,D}(t)+G_{R,A}(t)+G_{R,c}(t)$ and $G_{I,s}(t)=G_{I,D}(t)+G_{I,A}(t)+G_{I,c}(t)$ as follows:
\ben
k_{FG}=\frac{2\pi}{\hbar^2}|J|^2\int d\omega \int d\omega' I_A(\omega){\mathcal E}_D(\omega') I_c(\omega'-\omega) , \label{eq:kfg-spectral}
\een
where 
\ben
I_c(\omega)&=&\frac{1}{2\pi}\int dt \exp \Bigg \{ i\left (\omega-\frac{\lambda_c}{\hbar} \right )t  \nonumber \\
&&\hspace{.5in}-G_{R,c}(t)-iG_{I,c}(t) \Bigg \}  .  \label{eq:ic-exp}
\een
Note that Eq. (\ref{eq:kfg-spectral}) clearly demonstrates how common modes contribute to the deviation from F\"{o}rster's spectral overlap expression. Equation (\ref{eq:ic-exp})  is like the absorption lineshape except that the coupling reorganization energy $\lambda_c$ and the coupling lineshape functions $G_{R,c}(t)$ and $G_{I,c}(t)$ are used.  
Another interesting case is the strong coupling and high temperature limit where one can make Gaussian approximation, which however does not apply to the low temperature situation we are considering here.  

\begin{figure}
\includegraphics[width=3in]{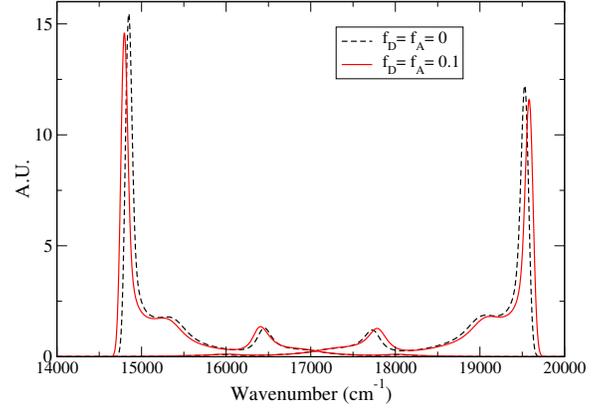}
\caption{\label{abs-em}Absorption lineshapes of \re{TDI-Ph} (left) and emission lineshapes of \re{Ph-PDI} (right) for two cases, $f_D=f_A=0$ and $f_D=f_A=0.1$.}
\end{figure}

We have evaluated the rate expression Eq. (\ref{eq:k-fg-common}) \re{at $1.4\ {\rm K}$}, directly employing the models of the bath spectral densities given by Eqs. (\ref{eq:jd})-(\ref{eq:jb}) and the parameters reported in Table \ref{table-bo}, but with additional approximations.  First, we ignored the contribution of other surrounding modes ($s_D=s_A=0$) that have not been \re{incorporated into} the Brownian oscillator model, assuming that they do not make significant contribution to the non-F\"{o}rster behavior.  Second, we considered only the first two Brownian oscillator terms ($k=1,2$) for ${\mathcal J}_{_{\rm Ph-PDI}}(\omega)$ and ${\mathcal J}_{_{\rm TDI-Ph}}(\omega)$.  This approximation is justified by the fact that the third Brownian oscillator terms have  much smaller magnitudes than others and that their frequencies are outside the regions where the donor and acceptor energies overlap most.  Figure \ref{abs-em} shows the absorption and emission lineshapes \re{also at $1.4\ {\rm K}$ for two different cases of parameters $f_D$ and $f_A$}, which all reproduce the overall qualitative features of experimental lineshapes \re{(see Fig. 10 of Ref. \onlinecite{hinze-jcp128})}   although finer details are missing.  It is important to note that the lineshapes in the overlap region are fairly insensitive to the two choices of $f_D=f_A=0$ and $0.1$ as shown.  If the F\"{o}rster's spectral overlap expression were correct, the two cases should result in almost the same rates.  However, this is not the case as will be shown below.

\begin{figure}
\includegraphics[width=3in]{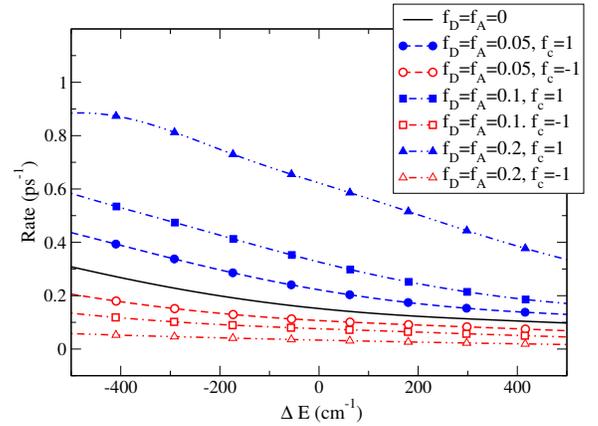}
\caption{ \label{rates} Exciton transfer rates calculated for different values of $f_D$, $f_A$, and $f_c$ versus $\Delta E=E_{D,em}-E_{A,ab}-\langle E_{D,em}\rangle+\langle E_{A,ab}\rangle$, where $\langle E_{D,em}\rangle = 2.3706\ {\rm eV} $ and $\langle E_{A,ab}\rangle = 1.9075\ {\rm eV}$ are used.    } 
\end{figure}

\re{Figure \ref{rates} provides series of theoretical rates calculated versus the relative energy bias $\Delta E$ between the emission energy of the donor $E_{D,em}$ and the absorption  energy of the acceptor $E_{A,sb}$ with respect to their average values $\langle E_{D,em}\rangle$ and $\langle E_{A,ab} \rangle$.  Experimental values  (see Table \ref{Excitation-energy}) were used for these average values. The range of $\Delta E$ chosen covers the range of disorder in this value due to the inhomogeneity of the environment around each molecule, as reported from single molecule spectroscopy experiments. For the donor-acceptor electronic coupling, we have used $94.1\ {\rm cm^{-1}}$.}  Four choices of $f_D=f_A=0$, $0.05$, $0.1$, and $0.2$ were considered.  For nonzero values of these parameters, two limits of positive ($f_c=1$) and negative ($f_c=-1$) contributions were considered.  For the bridge bath spectral density ${\mathcal J}_B(\omega)$, we have considered only the second Brownian oscillator ($k=2$), which is expected to make primary contribution to the rate.    

\re{The results in Fig.  \ref{rates} show that even small positive contributions of the bridge vibrational modes can cause significant enhancement of the rate, regardless of the variation of $\Delta E$ in the range shown.}  For example, already for the case of  $f_D=f_A=0.1$, the rates can be more than a factor of 2 larger than the rates without common modes.  On the other hand, Fig. \ref{abs-em} shows that the resulting change in lineshapes is very minor. \re{The mechanism of the enhancement of rates without affecting F\"{o}rster's spectral overlap can be understood clearly from Eq. \ref{eq:kfg-spectral}. The results provided in Fig. \ref{rates}} clearly illustrate that the contribution of common modes alone can explain the puzzling enhancement of rates seen\cite{metivier-prl98,hinze-jcp128} for {\bf D1}, \re{which was not fully explained by the enhancement of the effective electronic coupling constant alone.   }  

Another potentially important non-F\"{o}rster mechanism is the inelastic effect that can occur when the electronic coupling between the donor and the acceptor is significantly modulated by quantum vibrational degrees of freedom.  For \re{both {\bf D1} and {\bf D2}}, quantitatively reliable assessment of the inelastic effect requires additional set of calculations and simulations that go beyond the scope of this work.  For this reason, we here provide only a brief account of a simple model that can help gain some qualitative understanding. 

Due to near collinear arrangement of transition dipoles of donor and acceptor units, it is not expected that torsional \re{modes} make significant contributions. The vibration of the bond distance is not likely to have significant effect either because its amplitude is much smaller than \re{that of} the donor-acceptor distance.  This leaves the bending (or puckering motion) of the donor and acceptor units \re{as the most likely source for inelastic effects}.   While it is ideal to examine the dependence of the full electronic coupling on the bending angle, for simplicity, we consider only the dipole approximation.    Then, the corresponding electronic coupling, Eq. (\ref{eq:j-dp}), can be expressed  as
\begin{eqnarray}
J&=&\frac{\mu_{D}\mu_{A}}{n_{r}^{2}R^{3}}\left [\sin(\pi-\theta_{D})\sin \theta_{A}\cos(\phi_{D}-\phi_{A}) \right .\nonumber \\
&&\left . -2\cos(\pi- \theta_{D})\cos \theta_{A}\right ] \ ,  \label{eq:j-angle-1}
\end{eqnarray}
where $R$ is the distance between the centers of the two transition dipoles, $\theta_{D}$ and $\theta_{A}$ are axial angles, $\phi_{D}$ and $\phi_{A}$ are dihedral angles.  Let us  define $R_b$ as the length of the bridge unit of the dyad, $R_D$ the distance from the center of the electronic transition for Ph-PDI to the beginning part of the bridge unit and $R_A$ the distance from the ending part of the bridge unit to the center of the electronic transition for TDI-Ph. Then,  introducing the following two distances:
\ben
&&R_0=R_b+R_D+R_A  \ , \\
&&R_m=\frac{1}{2}(R_D+R_A) \ ,
\een
and making further simplification that $\phi_{D}=\phi_{A}$ and $\theta_{D}=\theta_{A}=\theta$, we can express  the distance $R$  as follows:
\be
R=R_b+R_D\cos \theta+R_A \cos \theta=R_0-2R_m(1-\cos \theta) \  .
\ee
Inserting the above expression into Eq. (\ref{eq:j-angle-1}) and assuming now $\hat \theta$ as a quantum observable, the electronic coupling operator can be expressed as 
\begin{equation}
\hat J=J_0\frac{1+\cos^2\hat \theta}{2[1-2r(1-\cos\hat \theta)]^3}\simeq J_{0}\{1+\sum_{k=1}^{4}p_{k}(1-\cos k\hat \theta)\}, \label{eq:j-mod}
\end{equation}
where $J_{0}$ is the electronic coupling when $\theta=0$, $p_{1}=21r/4+42r^{2}$, $p_{2}=-1/4-3r/2-18r^{2}$, $p_{3}=3r/4+6r^{2}$, and $p_{4}=-3r^{2}/2$, with $r=R_{m}/R_{0}$.  Under the assumption that the amplitude of $\theta$ is small and can be expressed as a linear combination of bath operators, namely,  $\hat \theta =\sum_n g_{nJ}(\hat b_{n}+\hat b_{n}^{\dagger})$, and introducing the corresponding bath spectral density, ${\mathcal J}_J(\omega)\equiv \pi \hbar \sum_{n} \delta (\omega-\omega_{n})\omega_{n}^2 g_{nJ}^2$, it is now possible to apply inelastic RET theories\cite{jang-jcp127,yang-jcp137} to the above model, Eq. (\ref{eq:j-mod}).  \re{We tried simple calculations assuming a super-Ohmic spectral density with cutoff frequency of $500\ {\rm cm^{-1}}$ for ${\mathcal J}_J(\omega)$, and found that the rate for {\bf D1} changes by about 10\% at most. While it is necessary to use more accurate  ${\mathcal J}_J(\omega)$ based on molecular dynamics simulations for a reliable assessment, we expect it would not result in much change in the contribution of the inelastic effect.  This is because even a very large value} of $\theta=20^o$ in Eq. (\ref{eq:j-mod}) does not affect the electronic coupling constant more than 5\% due to competing effects of the numerator and the denominator in Eq. (\ref{eq:j-mod}).  Thus,  while it remains a significant factor, we conlude that the inelastic effect is not the major source of non-FRET behavior for {\bf D1}. \vspace{.1in}\\

 \section{Conclusion}
 In this work, we have reported results of comprehensive computational study of {\bf D1} and {\bf D2}, mainly employing DFT and TD-DFT methods with B3LYP functional, and have also provided theoretical analysis of non-FRET mechanisms \re{for} {\bf D1}.  We found that the finite size effect of transition densities and the contribution of bridge units can account for the enhancement of the rate by about a factor of $4$ for {\bf D1} and about a factor of 3 for {\bf D2}, which are consistent with previous studies.  We then considered three \re{plausible} non-FRET rate mechanisms \re{for {\bf D1}}, with particular attention to the common mode effects.  
 
 Based on model calculations incorporating realistic information of the energetics and Franck-Condon factors of {\bf D1}, we have \re{provided} quantitative evidence for significant enhancement of the rate by common mode effects.  Our results show that \re{even fairly small percentage of the bridge vibrational modes participating in the common modes} can enhance the rate by about a factor of 2, while not affecting single molecule lineshapes significantly.  This occurs partly due to the large energy between the donor and the acceptor excitation energies, \re{for which the rate can be sensitive to small changes of common modes.  Thus, even small increase in the positive contribution of common modes} can enhance the transfer rate significantly.  On the other hand,  for {\bf D2}, such enhancement \re{is not likely to} play a major role because of longer length and more branched structure of the bridge unit.   
 
Our theoretical analyses provide a convincing theoretical explanation of the source of the discrepancy between experimental RET rates and the best theoretical estimates\cite{fuckel-jcp128,curutchet-jpcb112} available for {\bf D1}, which has \re{not been fully accounted for} more than a decade. \re{Our conclusion is that the enhancement of the electronic coupling due to delocalization of exciton to the bridge unit and the positive effect of common modes, combined together, can explain the discrepancy.}  \re{While we have reached this conclusion based on a comprehensive theoretical consideration}, further experimental and computational studies are \re{still} needed to confirm our findings here.  Comparison of experimental and computational rates for a series of similar systems but with different energetics between the donor and the acceptor and different vibrational structures for the bridge unit can help corroborate the conclusion of this work.

\acknowledgments
Seogjoo J. Jang (SJJ) thanks the support of the National Science Foundation  through CAREER award (CHE-0846899) during the initial stage, while Lei Yang (LY) was working at Queens College, and award No. CHE-1900170 for the completion of this project.   LY is supported by the National Natural Science Foundation of China (21503114), Nanjing University of Posts and Telecommunications Scientiﬁc Foundation NUPTSF (NY215056). 

\begin{center}
{\bf  Data Availability Statements} \\
\end{center}
Most data that support the findings of this article are contained in this article.  Additional data are available from the corresponding author upon reasonable request.

\noindent

\appendix

\section{Structural information on the ground electronic states}

 \begin{table}
\caption{\label{Selected-bond} Major bond distances (${\rm \AA}$) of Ph-PDI,TDI-Ph, {\bf D1} and {\bf D2} in the ground electronic state and the lowest excited state (in parentheses).  The ground state structure was optimized by DFT with B3LYP/6-31G(d) level, and the  excited state was optimized by CIS/6-31G(d)(except that the structure of {\bf D2} was optimized at the CIS/3-21G(d) level. \vspace{.01in}\\}
        \begin{tabular}{c|c|ccc}
            \hline
            \hline
        Unit   & Labels   & Monomer   & Dyad 1  & Dyad 2   \\
                        \hline
             Ph-PDI & $1-2$     & 1.394 (1.377)  & 1.411 (1.392)  & 1.408 (1.391)  \\
                         & $2-3$     & 1.222 (1.197)  & 1.220 (1.193)  & 1.222 (1.214)  \\
                         & $2-4$     & 1.483 (1.477)  & 1.484 (1.486)  & 1.483 (1.478)  \\
                        & $7-8$     & 1.396 (1.415)  & 1.396 (1.373)  & 1.396 (1.372)  \\
                        & $4-9$     & 1.383 (1.389)  & 1.384 (1.363)  & 1.384 (1.359)  \\
                       & $11-12$   & 1.430 (1.426)  & 1.430 (1.410)  & 1.430 (1.408)  \\
                        & $13-14$   & 1.384 (1.391)  & 1.384 (1.363)  & 1.384 (1.359)  \\
                        & $10-15$   & 1.396 (1.417)  & 1.396 (1.373)  & 1.396 (1.372)  \\        
         \hline

         TDI-Ph & $1-2$       & 1.395 (1.378)   & 1.412 (1.393)   & 1.408 (1.392)   \\
                       & $2-3$        & 1.222 (1.197)   & 1.221 (1.196)   & 1.223 (1.217)  \\
                       &$2-4$        & 1.481 (1.477)   & 1.482 (1.479)   & 1.481 (1.470)  \\
                       & $7-8$       & 1.399 (1.401)   & 1.399 (1.401)   & 1.399 (1.401)  \\
                       &$4-9$        & 1.385 (1.378)   & 1.386 (1.379)   & 1.385 (1.375)   \\
                       &$11-12$   & 1.439 (1.439)   & 1.439 (1.439)   & 1.439 (1.438)   \\
                        &$13-14$      & 1.394( 1.409)   & 1.394 (1.409)   & 1.394 (1.408)  \\
                        & $10-15$     & 1.394 (1.409)   & 1.394 (1.409)   & 1.394 (1.408)    \\
                         & $19-22$     & 1.482 (1.479)   & 1.481 (1.479)   & 1.482 (1.471)    \\
                          &$22-23$     & 1.223 (1.197)   & 1.223 (1.197)   & 1.223 (1.217)    \\
                          \hline
                          \hline
         \end{tabular}
        \end{table}

Some bond lengths of these compounds for both ground and excited states, which were determined by DFT and CIS calculations respectively, are listed in Table~\ref{Selected-bond}.  These data show that, in the ground states, most of the bond lengths of TDI-Ph or Ph-PDI molecular parts are almost the same except for the ${\rm N1-C2}$ bond (see Fig. \ref{Sketch-map}).
The differences in ${\rm N1-C2}$ bond lengths (i) between Ph-PDI and {\bf D1}, (ii) between Ph-PDI and {\bf D2},  (iii) between TDI-Ph and {\bf D1}, and (iv) between TDI-Ph and {\bf D2} are respectively (i) 0.017,  (ii) 0.014, (iii) 0.017 and (iv) 0.013 ${\rm \AA}$. For other bonds, the largest bond length difference is 0.004 ${\rm \AA}$.
The distance between the centers of PDI and TDI are 27.9 ${\rm \AA}$ and 40.2 ${\rm \AA}$ for {\bf D1} and {\bf D2}, respectively, which are comparable to the results obtained previously (28 and 41 ${\rm \AA}$, respectively).\cite{hinze-jcp128,hinze-jpca109}
Hinze $et~al.$\cite{hinze-jpca109} also measured the angle between the long axis of PDI (TDI) and center distance, which are 136$^{\circ}$ (164$^{\circ}$). These are similar to our calculated values 135.6$^{\circ}$ (162.5$^{\circ}$).

 For the excited states, TDI-Ph molecular parts are very similar in all the molecules.  Thus,  the ${\rm N1-C2}$ bond length difference of TDI part (0.015 ${\rm \AA}$) is the largest one between TDI-Ph and {\bf D1}.  Differences in all other bond lengths are smaller than 0.005 ${\rm \AA}$. For TDI-Ph and {\bf D2}, the largest bond length difference is 0.02${\rm\ \AA}$. These bond length differences indicate that the nature of electronic excitation in the TDI parts are similar for both {\bf D1} and {\bf D2}.   However, it is somewhat complicated for PDI molecular parts. Significant change in bond distance can be seen for ${\rm C7-C8}$, ${\rm C4-C9}$, ${\rm C11-C12}$, ${\rm C13-C14}$, and ${\rm C10-C15}$ bonds of the PDI part. The differences of these bonds between the excited state of Ph-PDI and those of two dyads are about 0.04 ${\rm \AA}$. On the other hand, the corresponding values  between the ground state of Ph-PDI and excited states of two dyads are only about 0.02 ${\rm \AA}$. These indicate that excitons in Ph-PDI portions of dyads are delocalized towards the bridge part to some extent.  This can have non-negligible effect on the mechanism of RET.

\begin{figure}
(a)\makebox[3in]{ }\\
\includegraphics[width=3.3in]{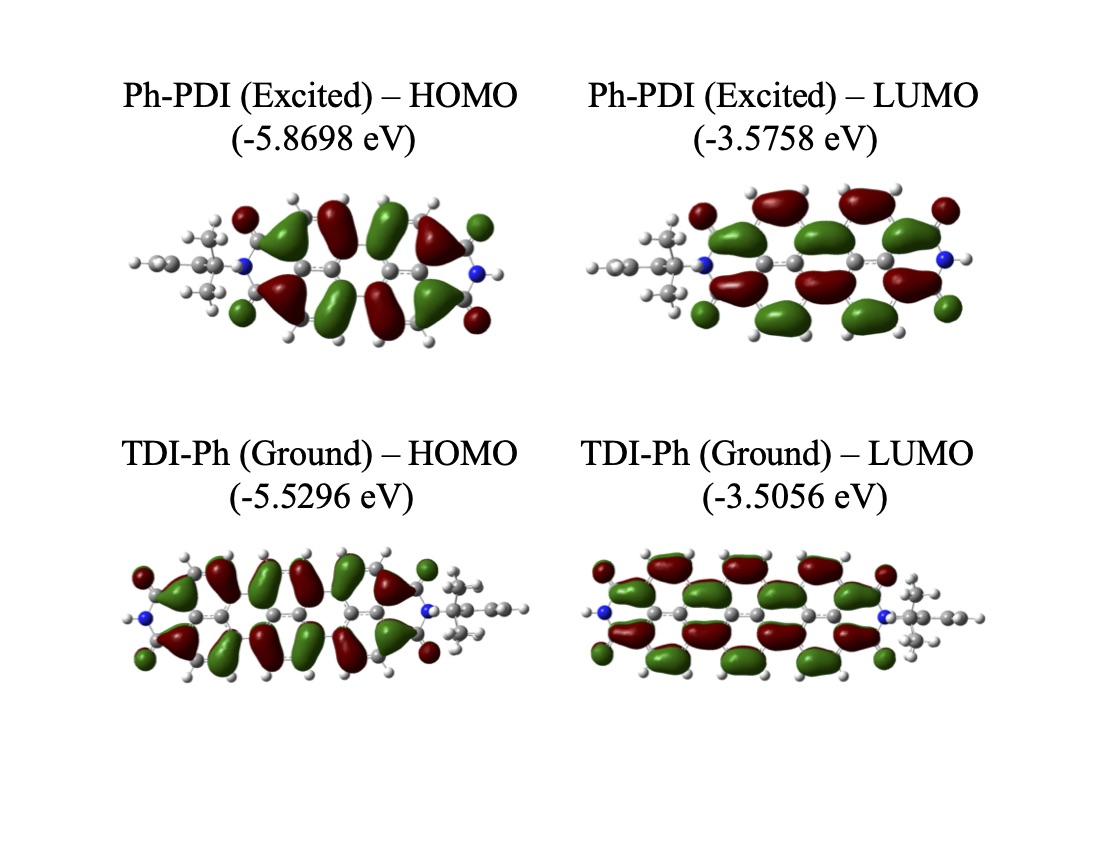} \vspace{-.5in}\\
(b)\makebox[3in]{ }\\
\includegraphics[width=3.3in]{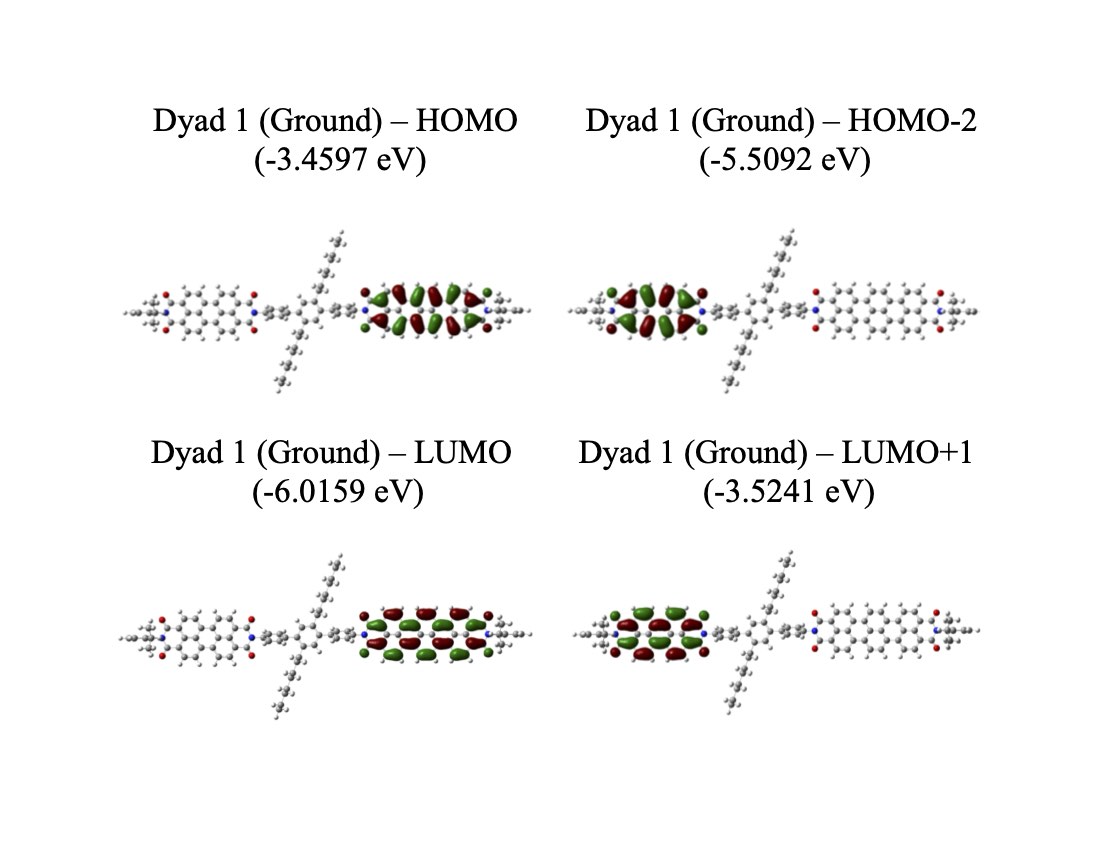} \vspace{-.5in}\\
(c)\makebox[3in]{ }\\
\includegraphics[width=3.3in]{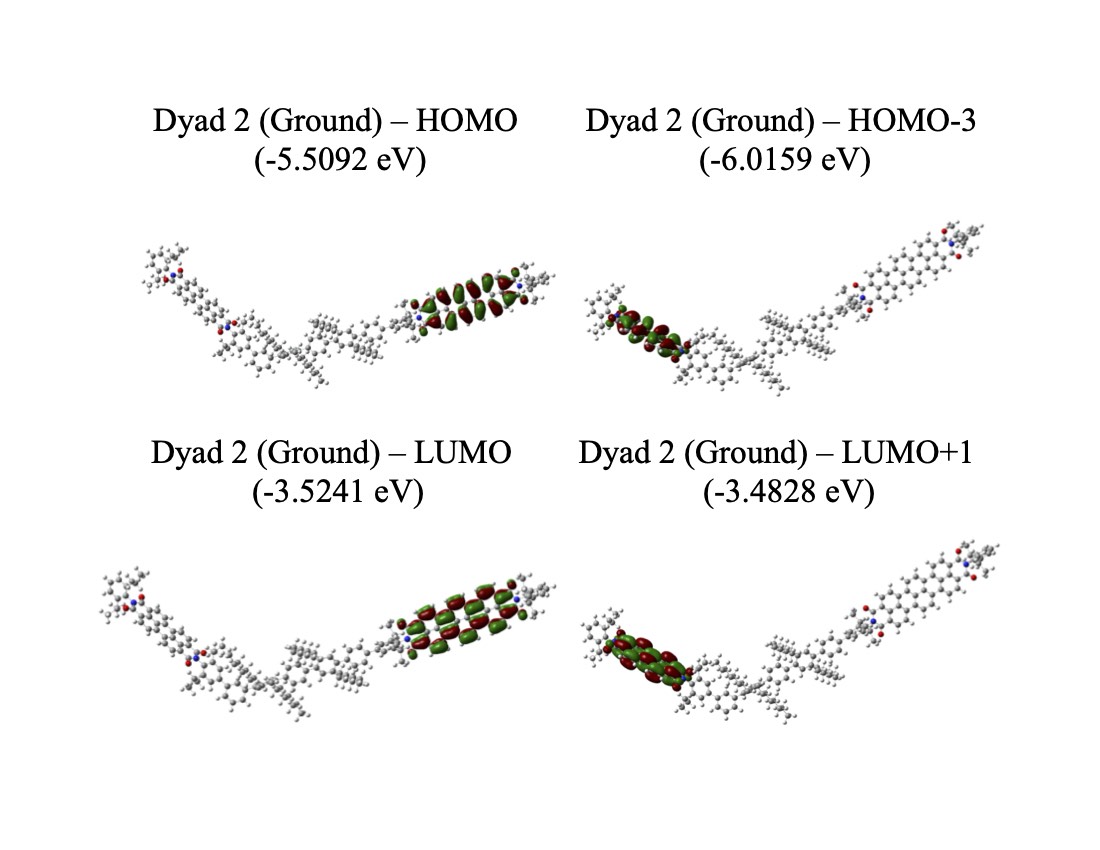} \\
\caption{Frontier molecular orbitals corresponding to the peak spectrum and their orbital energies. }
\label{Frontier-orbitals}
\end{figure}

Figure~\ref{Frontier-orbitals} shows the highest occupied molecular orbitals (HOMOs), the lowest unoccupied molecular orbitals (LUMOs), and other MOs that have major CI coefficients in the first excited state. All of these frontier orbitals have $\pi$ characteristics, and each of them is dominantly localized  at either Ph-PDI or TDI-Ph part.
For Ph-PDI and TDI-Ph, the lowest singlet excited state corresponds almost exclusively to the excitation from the HOMO to the LUMO of Ph-PDI and TDI-Ph, respectively.  This can be confirmed from changes of bond lengths upon excitations and relevant values for HOMO and LUMO orbitals. For instance, the HOMO of Ph-PDI has nodes across ${\rm C8-C9}$, ${\rm C14-C15}$ and ${\rm C7-C10}$ bonds, whereas the LUMO  has bonding characteristics in these regions. These bond lengths become shorter for the excited electronic state. On the other hand, ${\rm C4-C9}$, ${\rm C7-C8}$, ${\rm C10-C15}$ and ${\rm C13-C14}$ bonds become longer, which is consistent with the bonding characteristics of HOMO and the antibonding characteristics of LUMO in these regions. The same is true for TDI-Ph.

For {\bf D1} and {\bf D2}, there are two possible excited states resulting from linear combinations of those for monomers. As expected, the related HOMOs and LUMOs of these dyads are combinations of corresponding ones of individual Ph-PDI and TDI-Ph, respectively. For {\bf D1}, the HOMO and HOMO-2 are consistent with HOMOs of TDI-Ph and Ph-PDI, respectively.  For {\bf D2}, the HOMO and HOMO-3 are consistent with HOMOs of TDI-Ph and Ph-PDI, respectively.   On the other hand, for both dyads, the LUMO and LUMO+1 are consistent with LUMOs of TDI-Ph and Ph-PDI, respectively.


\begin{thebibliography}{10}

\bibitem{forster-ap}
T. F\"{o}rster, Ann. Phys. (Berlin) {\bf 437},  55  (1948).

\bibitem{forster-dfs}
{T. F\"{o}rster}, Discuss. Faraday Soc. {\bf 27},  7  (1959).

\bibitem{silbey-arpc27}
R. Silbey, Annu. Rev. Phys. Chem. {\bf 27},  203  (1976).

\bibitem{ret}
{D. L. Andrews and A. A. Demidov, Ed.}, {\em Resonance Energy Transfer} (John
  Wiley \& Sons, Chichester, 1999).

\bibitem{nitzan}
A. Nitzan, {\em Chemical Dynamics in Condensed Phases} (Oxford University
  Press, Oxford, 2006).

\bibitem{scholes-arpc54}
G.~D. Scholes, Annu. Rev. Phys. Chem. {\bf 54},  57  (2003).

\bibitem{olaya-castro-irpc30}
A. Olaya-Castro and G.~D. Scholes, Int. Rev. Phys. Chem. {\bf 30},  49  (2011).

\bibitem{jang-wires3}
S. Jang and Y.-C. Cheng, WIREs Comput. Mol. Sci. {\bf 3},  84  (2013).

\bibitem{bardeen-arpc65}
C.~R. Bardeen, Annu. Rev. Phys. Chem. {\bf 65},  127  (2014).

\bibitem{haacke-burghardt}
I. Burghardt and E. Stefan~Haacke, {\em Ultrafast Dynamics at the Nanoscale:
  Biomolecules and Supramolecular Assemblies} (Pan Stanford Publishing,
  Singapore, 2017).

\bibitem{jang-rmp90}
S.~J. Jang and B. Mennucci, Rev. Mod. Phys. {\bf 90},  035003  (2018).

\bibitem{stryer-pnas58}
L. Stryer and R.~P. Haugland, Proc. Natl. Acad. Sci. USA {\bf 58},  719
  (1967).

\bibitem{roy-nm5}
R. Roy, S. Hohng, and T. Ha, Nat. Methods {\bf 5},  507  (2008).

\bibitem{selvin-nsb7}
P.~R. Selvin, Nature Struct. Biol. {\bf 7},  730  (2000).

\bibitem{sahoo-jppc12}
H. Sahoo, J. Photochem. Photobio. C: Photochem. Rev. {\bf 12},  20  (2011).

\bibitem{heyduk-cob13}
T. Heyduk, Curr. Opin. Biotech. {\bf 13},  292  (2002).

\bibitem{schuler-cosb18}
B. Schuler and W. Eaton, Curr. Opin. Struct. Biol. {\bf 18},  16  (2008).

\bibitem{ha-pnas93}
T. Ha {\it et~al.}, Proc. Natl. Acad. Sci. {\bf 93},  6264  (1996).

\bibitem{weiss-science283}
S. Weiss, Science {\bf 283},  1676  (1999).

\bibitem{guo-pccp16}
Q. Guo, Y.~F. He, and H.~P. Lu, Phys. Chem. Chem. Phys. {\bf 16},  13052
  (2014).

\bibitem{basak-pccp16}
S. Basak and K. Chattopadhyay, Phys. Chem. Chem. Phys. {\bf 16},  11139
  (2014).

\bibitem{chung-pccp16}
H.~S. Chung and I.~V. Gopich, Phys. Chem. Chem. Phys. {\bf 16},  18644  (2014).

\bibitem{stockmar-jpcb120}
F. Stockmar, A.~Y. Kobitski, and G.~U. Nienhaus, J. Phys. Chem. B {\bf 120},
  641  (2016).

\bibitem{beljonne-jpcb113}
D. Beljonne, C. Curutchet, G.~D. Scholes, and R.~J. Silbey, J. Phys. Chem. B
  {\bf 113},  6583  (2009).

\bibitem{schuler-pnas102}
B. Schuler {\it et~al.}, Proc. Natl. Acad. Sci., USA {\bf 102},  2754  (2005).

\bibitem{langhals-jacs132}
H. Langhals {\it et~al.}, J. Am. Chem. Soc. {\bf 132},  16777  (2010).

\bibitem{dayal-jacs128}
S. Dayal {\it et~al.}, J. Am. Chem. Soc. {\bf 128},  13974  (2006).

\bibitem{metivier-prl98}
R. M\'{e}tivier, F. Nolde, K. M\"{u}llen, and T. Basch\'{e}, Phys. Rev. Lett.
  {\bf 98},  047802  (2007).

\bibitem{hinze-jcp128}
G. Hinze {\it et~al.}, J. Chem. Phys. {\bf 128},  124516  (2008).

\bibitem{athanasopoulos-jpcl8}
S. Athanasopoulos {\it et~al.}, J. Phys. Chem. Lett. {\bf 8},  1688  (2017).

\bibitem{sumi-prl50}
H. Sumi, Phys. Rev. Lett. {\bf 50},  1709  (1983).

\bibitem{sumi-jpcb103}
H. Sumi, J. Phys. Chem. B {\bf 103},  252  (1999).

\bibitem{jang-prl92}
S. Jang, M.~D. Newton, and R.~J. Silbey, Phys. Rev. Lett. {\bf 92},  218301
  (2004).

\bibitem{jang-cp275}
S. Jang, Y.~J. Jung, and R.~J. Silbey, Chem. Phys. {\bf 275},  319  (2002).

\bibitem{jang-jcp127}
S. Jang, J. Chem. Phys., {\bf 127},  174710  (2007).

\bibitem{jang-jcp129}
S. Jang, Y.-C. Cheng, D.~R. Reichman, and J.~D. Eaves, J. Chem. Phys. {\bf
  129},  101104  (2008).

\bibitem{jang-jcp131}
S. Jang, J. Chem. Phys. {\bf 131},  164101  (2009).

\bibitem{jang-jcp135}
S. Jang, J. Chem. Phys. {\bf 135},  034105  (2011).

\bibitem{yang-jcp137}
L. Yang, M. Devi, and S. Jang, J. Chem. Phys. {\bf 137},  024101  (2012).

\bibitem{jang-wires}
S. Jang, H. Hossein-Nejad, and G. D. Scholes. in {\it Quantum Effects in Biology}, edited by M. Mohseni, Y.
Omar, G. Engel, and M. Plenio (Cambridge University Press, Cambridge, 2014)

\bibitem{jang-prl113}
S. Jang, S. Hoyer, G.~R. Fleming, and K.~B. Whaley, Phys. Rev. Lett. {\bf 113},
   188102  (2014).

\bibitem{jang-jpcc123}
S.~J. Jang, J. Phys. Chem. C {\bf 123},  5767  (2019).

\bibitem{hennebicq-jcp130}
E. Hennebicq {\it et~al.}, J. Chem. Phys. {\bf 130},  214505  (2009).

\bibitem{du-cs9}
M. Du {\it et~al.}, Chem. Sci. {\bf 9},  6659  (2018).

\bibitem{jang-exciton}
S.~J. Jang, {\em Dynamics of Molecular Excitons (Nanophotonics Series)}
  (Elsevier, Amsterdam, 2020).

\bibitem{fuckel-jcp128}
B. F\"{u}ckel {\it et~al.}, J. Chem. Phys. {\bf 128},  074505  (2008).

\bibitem{curutchet-jpcb112}
C. Curutchet, B. Mennucci, G.~D. Scholes, and D. Beljonne, J. Phys. Chem. B
  {\bf 112},  3759  (2008).

\bibitem{hubner-jcp120}
C.~G. H\"{u}bner {\it et~al.}, J. Chem. Phys. {\bf 120},  10867  (2004).

\bibitem{hinze-jpca109}
G. Hinze {\it et~al.}, J. Phys. Chem. A {\bf 109},  6725  (2005).

\bibitem{fuckel-jcp125}
B. F\"{u}ckel {\it et~al.}, J. Chem. Phys. {\bf 125},  144903  (2006).

\bibitem{kohn-pr140}
W. Kohn and L.~J. Sham, Phys. Rev. {\bf 140},  A1133  (1965).

\bibitem{lee-prb37}
C. Lee, W. Yang, and R.~G. Parr, Phys. Rev. B {\bf 37},  785  (1988).

\bibitem{yanai-cpl393}
T. Yanai, D. Tew, and N. Handy, Chem. Phys. Lett. {\bf 393},  51  (2004).

\bibitem{zhao-tca120}
Y. Zhao and D.~G. Truhlar, Theo. Chem. Acc. {\bf 120},  215  (2008).

\bibitem{runge-prl52}
E. Runge and E.~K.~U. Gross, Phys. Rev. Lett. {\bf 52},  997  (1984).

\bibitem{foresman-jpc96}
{J. B. Foresman, M. Head-Gordon, J. A. Pople, and M. J. Frisch}, J. Phys. Chem.
  {\bf 96},  135  (1992).

\bibitem{g09}
M.~J. Frisch {\it et~al.}, Gaussian~09 {R}evision {A}.1, gaussian Inc.
  Wallingford CT 2009.

\bibitem{krueger-jpcb102}
B.~P. Krueger, G.~D. Scholes, and G.~R. Fleming, J. Phys. Chem. B {\bf 102},
  5378  (1998).

\bibitem{caprasecca-jctc8}
S. Caprasecca, C. Curutchet, and B. Mennucci, J. Chem. Theory Comput. {\bf 8},
  4462  (2012).

\bibitem{chen-jcp129}
H.-C. Chen, Z.-Q. You, and C.-P. Hsu, J. Chem. Phys. {\bf 129},  084708
  (2008).

\bibitem{yang-jacs132}
L. Yang, S. Caprasecca, B. Mennucci, and S. Jang, J. Am. Chem. Soc. {\bf 132},
  16911  (2010).

\bibitem{jang-njp15}
S. Jang, T. Berkelbach, and D.~R. Reichman, New J. Phys. {\bf 15},  105020
  (2013).

\bibitem{holstein-ap8-1}
T. Holstein, Ann. Phys. {\bf 8},  325  (1959).

\bibitem{holstein-ap8-2}
T. Holstein, Ann. Phys. {\bf 8},  343  (1959).

\bibitem{rackovsky-mp25}
S. Rackovsky and R. Silbey, Mol. Phys. {\bf 25},  61  (1973).

\bibitem{tong-jcp153}
Z. Tong {\it et~al.}, J. Chem. Phs. {\bf 153},  044105  (2020).

\bibitem{reimers-jcp115}
J.~R. Reimers, J. Chem. Phys. {\bf 115},  9103  (2001).

\end{thebibliography}
\end{document}